\documentclass{article}

\usepackage{amscd}
\usepackage{amsmath,amssymb}

\newcommand{\zstar}{z^{\ast}}
\newcommand{\ystar}{y^{\ast}}
\newcommand{\tf}{\tilde{f}}
\newcommand{\bB}{{\mathbf B}}
\newcommand{\bC}{{\mathbf C}}
\newcommand{\bT}{{\mathbf T}}
\newcommand{\val}{\nu}
\newcommand{\barS}{\overline{S}}

\newcommand{\al}{\alpha}
\newcommand{\gam}{\gamma}
\newcommand{\Gam}{\Gamma}
\newcommand{\Lam}{\Lambda}
\newcommand{\Lamc}{\Lambda^c}
\newcommand{\lam}{\lambda}
\newcommand{\kap}{\kappa}
\newcommand{\del}{\delta}

\newcommand{\Om}{\Omega}
\newcommand{\om}{\omega}
\newcommand{\sig}{\sigma}

\newcommand\mut[1]{\ignorespaces}

\DeclareMathOperator{\res}{res}
\DeclareMathOperator{\ress}{res}
\DeclareMathOperator{\Trr}{Tr}
\DeclareMathOperator{\ev}{ev}

\usepackage{amsthm}
\usepackage{graphics}
\newtheorem{example}{Example}
\newtheorem{lemma}{Lemma}
\newtheorem{theorem}{Theorem}
\newtheorem{remark}{Remark}
\newtheorem{proposition}{Proposition}
\newtheorem*{definition}{Definition}

\newcommand\F{{\mathbb F}}
\newcommand\N{{\mathbb N}}
\newcommand\Z{{\mathbb Z}}
\newcommand\Cc{{\mathcal C}}

\title{The Key Equation for One-Point Codes}

\date{July 23, 2008}

\author{Michael E. O'Sullivan\thanks{San Diego State University, {\tt mosullivan@sdsu.edu}.}, Maria Bras-Amor\'os\thanks{Universitat Rovira i Virgili, {\tt maria.bras@urv.cat}.}}

\begin{document}

\maketitle

\begin{abstract}
For Reed-Solomon codes, the key equation relates the syndrome
polynomial ---computed from the parity check matrix and the received
vector--- to two unknown polynomials, the locator and the evaluator.
The roots of the locator polynomial identify the error positions.
The evaluator polynomial, along with the
derivative of the locator polynomial, gives  the error values via 
the Forney formula.
The Berlekamp-Massey algorithm  efficiently computes the two
unknown polynomials.

This chapter shows how the key equation, the Berlekamp-Massey
algorithm, the Forney formula, and another formula for error
evaluation due to Horiguchi all generalize in a natural way to 
one-point codes.
The algorithm presented here is based on K\"otter's
adaptation of  Sakata's algorithm. 

\end{abstract}

\vskip2cm

{\small Published as {\bf M. E. O’Sullivan, M. Bras-Amorós, The Key Equation for One-Point Codes, Chapter 3 of Advances in Algebraic Geometry Codes, World Scientific, E. Martínez-Moro, C. Munuera, D. Ruano (eds.), vol. 5, pp. 99-152, 2008. ISBN 978-981-279-400-0.}}

\newpage

\tableofcontents

\newpage

\section{Introduction}

For Reed-Solomon codes, the key equation relates the syndrome
polynomial---computed from the parity check matrix and the received
vector---to two unknown polynomials, the locator and the evaluator.
The exact formulation of the key equation has evolved since
Berlekamp's introduction of the term~\cite{Berlekamp:Book}.  There are
also key equations for other algorithms, such as   
Sugiyama et al~\cite{sugiyama:key}, and 
Berlekamp-Welch~\cite{WelchBerl:patent}. 
The goal of this chapter is to show that  the key equation, the
Berlekamp-Massey algorithm and the error evaluation formulas of Forney
and Horiguchi~\cite{horiguchi}
all generalize  to one-point codes.  
An important aspect of the generalization is to treat the ideal of
error locator polynomials as a module over a polynomial ring in one
variable, which is essentially the approach K\"otter used in his
version  of the Berlekamp-Massey-Sakata algorithm~\cite{kot98}.
The chapter is divided into three main sections, Reed-Solomon codes,
Hermitian codes, and one-point codes.  
We have attempted to make each section as self-contained as possible, 
and to minimize the mathematical background required.

The section on Reed-Solomon codes gives a concise treatment of the
key-equation, the Berlekamp-Massey algorithm, 
and the error evaluation formulas in a manner that will generalize easily to
one-point codes.
Two aspects of our approach are atypical, though certainly not new.  
First, the locator polynomial vanishes at the 
error positions---as opposed to the usual definition which uses the
reciprocals of the positions---because this is more natural in the
context of algebraic geometry codes.   
Second, the syndrome is a rational polynomial---rather than a
polynomial---because this is in accord with the duality of codes on
algebraic curves.  
Theorem~\ref{t:BM} gives a very formal statement of the properties
satisfied by the intermediate polynomials computed in the
Berlekamp-Massey algorithm.  
Analogous results are established in the later sections for Hermitian
and one-point codes.
At the end of the section on Reed-Solomon codes 
we  briefly discuss the usual formulation
of the key equation---see for
example~\cite{Roth:Book,Blahut:Book}---and the 
connections with the 
Euclidean algorithm and  the algorithm of Sugiyama et al.  
There are also interesting connections to the Berlekamp-Welch
algorithm and to the list decoding algorithm of Lee and
O'Sullivan~\cite{LeeOS:RS}, but these are not developed here.

The section on Hermitian codes requires little if any background in
algebraic geometry, and only minimal familiarity with the algebra of
polynomial rings and Gr\"obner bases.
The presentation of this section closely parallels that of the section on
Reed-Solomon codes, so that overall similarity between the two
as well as the new complexities are as clear as possible.
The locator polynomial is replaced with the ideal of polynomials
vanishing at the error locations, and the problem is to find
several locator polynomials of minimal degree, one for each congruence
class modulo $q$, where the field size is $q^2$.  
The syndrome is again a rational polynomial, and the property of a
locator is that its product with the syndrome eliminates the
denominator, giving a  polynomial.  The product of the locator and the
syndrome also  may be used for error evaluation.  
K\"otter's  algorithm is essentially $q$ Berlekamp-Massey algorithms
operating in parallel, and the only place in which the algebra of the
curve is used is in the computation of recursions of candidate locator
polynomials with the syndrome.  
The Forney formula and Horiguchi formula for error evaluation are 
simple, but not obvious, generalizations of those for Reed-Solomon codes.

The section on one-point codes shows that the decoding algorithms and
formulas for Hermitian codes need only minor modification to apply to
general one-point codes.  The focus of this section is not
reproving the decoding results in the more general setting; 
instead, it is to establish the algebraic structure
that makes the algorithms work.  
In particular, we will need to use differentials, residues of
differentials, and  duality with respect to the residue map.
 This section does require the theory of
curves and algebraic function fields, but we have tried to build the
exposition using a small number of key results as a base.
The treatment is based on O'Sullivan~\cite{OS:key,OS:kot}, with, we
hope, improvements in exposition.  

\section{The key equation for Reed-Solomon codes}
In this section, we briefly discuss Reed-Solomon codes,  set up the
decoding problem and introduce the locator and evaluator polynomials.   
The syndrome is defined as  a rational polynomial, but it may also be seen
as a power series.  
We then present the key equation and the Berlekamp-Massey algorithm in 
a form that we will generalize  to codes from algebraic curves.  
We derive Horiguchi's  formula for error evaluation, which removes the need to compute the error evaluator polynomial.  Finally, we explore connections with the Euclidean algorithm. 

\subsection{Reed-Solomon codes}
\label{s:RS}
Let ${\F}_q$ be the  finite field of $q$ elements.
Given $n$ different elements $\alpha_1,\dots,\alpha_{n}$ of ${\mathbb F}_q$,
define the map $\ev:{\mathbb F}_q[x]\rightarrow {\mathbb F}_q^n$, $f\mapsto
(f(\alpha_1),\dots,f(\alpha_{n}))$.
The {\em generalized Reed-Solomon code} $GRS(\bar{\alpha}, k)$, 
where $\bar{\alpha} = (\alpha_1,\dots, \alpha_n)$, 
is defined as the image by
$\ev$ of the polynomials
in ${\mathbb F}_q[x]$ with degree at most $k-1$. It has generator matrix
$$\left(\begin{array}
{cccc}
1 & 1 &\dots& 1\\
\alpha_1 & \alpha_2 & \dots & \alpha_{n}\\
\alpha_1^2 & \alpha_2^2 & \dots & \alpha_{n}^2\\
\vdots & \vdots & \vdots &  \vdots \\
\alpha_1^{k-1} & \alpha_2^{k-1} & \dots & \alpha_{n}^{k-1}\\
\end{array} \right) $$

It is well known (see for instance \cite[\S 5.1]{Roth:Book})
that the parity check matrix of $GRS(\bar{\alpha},k)$ is then
$$\left(\begin{array}
{cccc}
1 & 1 &\dots& 1\\
\alpha_1 & \alpha_2 & \dots & \alpha_{n}\\
\alpha_1^2 & \alpha_2^2 & \dots & \alpha_{n}^2\\
\vdots & \vdots & \vdots &  \vdots \\
\alpha_1^{n-k-1} & \alpha_2^{n-k-1} & \dots & \alpha_{n}^{n-k-1}\\
\end{array}\right)\left(\begin{array}{ccccc}
\beta_1 & 0 & \dots & \dots & 0\\
0 & \beta_2 & 0 & \ddots & \vdots\\
\vdots & 0 & \ddots & \ddots& \vdots\\
\vdots & \ddots & 0 & \ddots& 0\\
0 & \dots & \dots & 0 & \beta_{n}\\
\end{array}\right)$$
for some $\beta_1,\beta_2,\dots,\beta_{n}\in{\mathbb F}_q$.
That is, $(c_1,c_2,\dots,c_{n})$ is in
$GRS(\bar{\alpha},k)$ if and only if
$(c_1\beta_1,c_2\beta_2,\dots,c_{n}\beta_{n})$
is in $GRS^{\perp}(\bar{\alpha}, n-k)$.

If the field size is $q$ and $n=q-1$ then it is said to be a
{\em conventional Reed-Solomon code} or just Reed-Solomon code
and we denote it by $RS(k)$.
In this case it can be proven that $\beta_i=\alpha_i$. So
the parity check matrix is
$$\left(\begin{array}
{ccccc}
\alpha_1 & \alpha_2 & \dots & \alpha_{n}\\
\alpha_1^2 & \alpha_2^2 & \dots & \alpha_{n}^2\\
\alpha_1^3 & \alpha_2^3 & \dots & \alpha_{n}^3\\
\vdots & \vdots & \vdots & \vdots \\
\alpha_1^{n-k} & \alpha_2^{n-k} & \dots & \alpha_{n}^{n-k}\\
\end{array}\right).$$

\subsection{Polynomials for decoding}
Suppose that a word $c\in GRS^{\perp}(\bar{\alpha},n-k)$ 
is transmitted and that the vector $u$ is received.
The vector  $e= u-c$ is the error vector.  We assume that $e$ has 
$t\leq\frac{n-k}{2}$ non-zero positions.
We will use  $c$, $u$,
 $e$ and $t$ throughout  this section.
The decoding task is to recover $e$ from $u$ and thereby get $c=u-e$.

We define the {\em error locator polynomial} associated to $e$ as
$$f^e=\prod_{j:e_j\neq 0}(x-\alpha_j)$$
and the {\em error evaluator polynomial} as
$$\varphi^e=\sum_{j:e_j\neq 0}e_j\prod_{\substack{k:e_k\neq 0 \\k\neq j }}(x-\alpha_k).$$

The utility of  the error locator polynomial and the error
evaluator polynomial is that the error
positions can be identified as the indices $j$ such that
$f^e(\alpha_j)=0$ and the error values
can be computed by the so-called Forney formula
given in the next lemma, whose verification is straightforward.

\begin{lemma}
\label{lemma:errorvalues}
If $e_j\neq 0$ then
$e_j=\frac{\varphi^e(\alpha_{j})}{{f^e}'(\alpha_{j})}.$
\end{lemma}

Another useful fact about $f^e$ and $\varphi^e$ is that
from the received vector we know the first coefficients of the power series in
$\frac{1}{x}$ obtained when dividing $\varphi^e$ by $f^e$.
This is shown in the next lemma.

\begin{lemma}
\label{lemma:RSsyn}
$\frac{\varphi^e}{f^e}=\frac{1}{x}\left(s_0+\frac{s_1}{x}+\frac{s_2}{x^2}+\cdots\right)$, where
$s_a=\sum_{j=1}^{n}e_j\alpha_j^a$.
In particular, for $a\leq n-k-1$,
$s_a=\sum_{j=1}^{n}u_j\alpha_j^a$.
\end{lemma}

\begin{proof}
\begin{eqnarray*}
\frac{\varphi^e}{f^e}=\sum_{j=1}^{n}\frac{e_j}{x-\alpha_j}&=&\frac{1}{x}\sum_{j=1}^{n}\frac{e_j}{1-\frac{\alpha_j}{x}}\\
&=&\frac{1}{x}\sum_{j=1}^{n}e_j\sum_{a=0}^\infty\left(\frac{\alpha_j}{x}\right)^a\\
&=& \frac{1}{x}\sum_{a=0}^\infty\frac{1}{x^a}\sum_{j=1}^{n}e_j\alpha_j^a\\
&=&\frac{1}{x}\sum_{a=0}^\infty\frac{s_a}{x^a}\\
\end{eqnarray*}
Looking at the parity check matrix of $GRS(\bar{\alpha},n-k)^\perp$,
it can be deduced that
for $a\leq n-k-1$,
$\sum_{j=1}^{n}c_j\alpha_j^a=0$. Hence,
$s_a=\sum_{j=1}^{n}e_j\alpha_j^a=\sum_{j=1}^{n}(u_j-c_j)\alpha_j^a=\sum_{j=1}^{n}u_j\alpha_j^a$.
\end{proof}

\begin{definition}
For a vector $e$, the {\em syndrome} of $e$ is $S=\frac{\varphi^e}{f^e}$.
The {\it syndrome of order } $a$  is $s_a= \sum_{j=1}^{n}u_j\alpha_j^a$.
\end{definition}

Just as any element of the field $\F_q(x)$ may be written as a Laurent
series in $x$, any $h \in \F_q(x)$ also may be written as a Laurent
series in $1/x$, $h = \sum_{a \leq d} h_ax^a$ for some $d \in \Z$.
If $h_d$ is nonzero in this expression, we say the {\em degree} of $h$ is
$d$, and if $h_d=1$ we say that $h$ is {\em monic}. 
Notice that $h \in \F_q[x]$ if and only if $h_a = 0$ for all
$a<0$ and that our definition of degree coincides with the usual one
on $\F_q[x]$. 
Henceforth, we will not use the form for $h$ given above.  Instead
we will write Laurent series in $1/x$ in the form
$h= \frac{1}{x} \sum_a h_a x^{-a}$.  It is understood that the sum is
over all integers $a \geq -d-1$  where $d$ is the degree of $h$.
In  this form, $h$ is a polynomial when $h_a=0$ for all $a \geq 0$.
As an example, the syndrome is $S = \frac{1}{x} \sum_{a\geq 0} s_a
x^{-a}$.  Its degree is $-1$, unless $s_0=0$.

\begin{lemma}
Let $f$ be a polynomial and let $\alpha\in{\mathbb F}_q$.
If the Laurent series in $\frac{1}{x}$
given by
$\frac{f}{x-\alpha}$
has no term of degree $-1$
then $f(\alpha)=0$.
\end{lemma}

\begin{proof}
There exists $g\in{\mathbb F}_{q}[x]$ such that
$f(x)=f(\alpha)+(x-\alpha)g(x)$. Then
\begin{eqnarray*}\frac{f(x)}{x-\alpha}&=&\frac{f(\alpha)}{x-\alpha}+g(x)\\
&=&g(x)+\frac{f(\alpha)}{x}\left(1+\frac{\alpha}{x}+ \left(\frac{\alpha}{x}\right)^2+\cdots\right)\\
&=&g(x)+\frac{f(\alpha)}{x}+\frac{\alpha f(\alpha)}{x^2}+\frac{\alpha^2 f(\alpha)}{x^3}+\cdots\\
\end{eqnarray*}
If the term of degree $-1$ is zero, then $f(\alpha)=0$.
\end{proof}

\begin{proposition}
\label{p:locator characterization}
If $f S$ has no terms of degrees $-1,-2,\dots,-t$ then
$f$ is a multiple of $f^e$.
In particular, if $fS $ is a polynomial then $f$ is a multiple of~$f^e$.
\end{proposition}

\begin{proof}
Suppose $f S$ has no terms of degrees
$-1,-2,\dots,-t$.
Suppose $e_j\neq 0$ and let
$$g(x)=\prod_{\substack{k:e_k\neq 0 \\k\neq j }}  (x-\alpha_k).$$
Note that $\deg{g}=t-1$ and so $fgS$ has no term of degree $-1$.
Now,
\begin{eqnarray*}
fgS&=&\sum_{k: e_k\neq0}\frac{e_kfg}{x-\alpha_k}\\
&=&e_j\frac{fg}{x-\alpha_j}+
\sum_{\substack{k:e_k\neq 0 \\k\neq j }}e_kf\frac{g}{x-\alpha_k}.\\
\end{eqnarray*}
Since $fgS$ has no term of degree $-1$
and the right term in the previous sum is a polynomial,
we deduce that $\frac{fg}{x-\alpha_j}$
has no term of degree $-1$. By the previous lemma,
$x-\alpha_j$ must divide $f$.
Since $j$ was chosen arbitrarily such that $e_j\neq 0$,
we conclude that $f^e$ must divide $f$.
\end{proof}

\subsection{The key equation and the Berlekamp-Massey algorithm}
We now present the version of the Berlekamp-Massey algorithm that will
be our model for generalization to codes from algebraic curves.
The Berlekamp-Massey algorithm finds the minimal solution to the key
equation.

\begin{definition}
We will say that polynomials $f, \varphi$ satisfy the {\em key equation} for syndrome $S$
when  $fS=\varphi$.
\end{definition}

\medskip

\begin{center}
{\bf The Berlekamp-Massey Algorithm}
\end{center}

\noindent {\bf Initialize:}
$\left(\begin{array}{cc}
f^{(0)}&\varphi^{(0)}\\g^{(0)}&\psi^{(0)}\\
\end{array}\right)=
\left(\begin{array}{cc}1&0\\0&-1\\
\end{array}\right)$
\medskip

\noindent {\bf Algorithm:} For $m=0$ to $n-k-1$,

\begin{quote}
$d=\deg{f^{(m)}}$

\medskip

$\mu=\sum_{a=0}^{d} f^{(m)}_{a}s_{a+(m-d)}$

\medskip 
$p = 2d-m-1$
\medskip

$U^{(m)}=\left\{\begin{array}{ll}
\left(\begin{array}{cc}
1 & -\mu x^p\\
0 & 1\\
\end{array}\right)& \mbox{ if }\mu= 0 \mbox{ or } p \geq 0
\\
\left(\begin{array}{cc}
x^{-p}&-\mu\\
1/\mu& 0
\end{array}\right)
& \mbox{ otherwise.}
\\
\end{array}
\right.$

\medskip

$\left(\begin{array}{cc}
f^{(m+1)}&\varphi^{(m+1)}\\g^{(m+1)}&\psi^{(m+1)}\\
\end{array}\right)
=U^{(m)}\left(\begin{array}{cc}
f^{(m)}&\varphi^{(m)}\\g^{(m)}&\psi^{(m)}\\
\end{array}\right)$

\end{quote}
\noindent {\bf Output:} $f^{(n-k)}$, $\varphi^{(n-k)}$.

Notice that this algorithm
uses only the syndromes of order
up to $n-k-1$
and these are exactly the syndromes that can be computed from the received vector.
We may think of $f^{(m)}, \varphi^{(m)}$ and also $g^{(m)}, \psi^{(m)}$
as  approximate solutions of the key equation.  The algorithm takes a
linear combination of two approximate solutions to create a better
approximation.

\begin{theorem}
\label{t:BM}
For all $m\geq 0$,
\begin{enumerate}
\item\label{it1}
$f^{(m)}$ is monic of degree at most $m$.
\item\label{it2}
$\deg{(f^{(m)}S-\varphi^{(m)})}\leq -m+\deg{f^{(m)}}-1$. In particular, $f^{(m)}S$
has no terms in degrees $-1,-2,\dots,-m+\deg{f^{(m)}}$.
\item\label{it3}
$g^{(m)}S-\psi^{(m)}$ is monic of degree
$-\deg{f^{(m)}}$.
\item\label{it4}
$\deg{(g^{(m)})}\leq m-\deg{f^{(m)}}$.
\end{enumerate}
\end{theorem}

\begin{proof}
We will proceed by induction on $m$.
It is easy to verify the case $m=0$.
Assume the statements are satisfied at step $m$.
Let $d = \deg {f^{(m)}}$.  Notice that $d \leq m$ by item~(\ref{it1}),
and $\mu$ is the coefficient of $x^{d-m-1}$ in
$f^{(m)}S$.  Furthermore, since $d-m-1<0$,
and $\varphi^{(m)}$ is a polynomial, $\mu $ is the coefficient of
$x^{d-m-1}$ in $f^{(m)}S-\varphi^{(m)}$.

If $\mu=0$, then the algorithm 
retains the polynomials from the $m$th iteration, {\it e.g.} 
$f^{(m+1)}=f^{(m)}$. 
The induction hypothesis immediately gives
items~(\ref{it1}),~(\ref{it3}), and~(\ref{it4}) of the theorem, and
item~(\ref{it2}) follows from $\mu=0$.

Consider the case when $p=2d-m-1\geq 0$ and $\mu\not=0$.
The algorithm sets $f^{(m+1)}=f^{(m)}-\mu x^{p}g^{(m)}$.
By the induction hypothesis,
\[\deg{( x^{p}g^{(m)})} \leq 2d-m-1+m-d = d-1,
\]
so $\deg(f^{(m+1)}) = \deg(f^{(m)})=d<m$ and
$f^{(m+1)}$
is monic, so item~(\ref{it1}) holds.
Now,
\[f^{(m+1)}S-\varphi^{(m+1)}=
(f^{(m)}S-\varphi^{(m)})-\mu
x^{p}(g^{(m)}S-\psi^{(m)}).\]
The degree of each term is $d-m-1$ and the coefficients of $x^{d-m-1}$
cancel. Thus 
$\deg( f^{(m+1)}S-\varphi^{(m+1)}) \leq -(m+1) + \deg( f^{(m+1)}) -1$,
as required. This proves item~(\ref{it2}).
Items~(\ref{it3}) and~(\ref{it4}) are trivial in this case, since
$g^{(m+1)}= g^{(m)}$ and $\varphi^{(m+1)}=\varphi^{(m)}$.

Finally, consider the case when $p=2d-m-1 <0$, in which
$f^{(m+1)}=x^{-p}f^{(m)}-\mu g^{(m)}$.
By computing the degrees of each summand, one can see that
$f^{(m+1)}$ is monic of degree $ m+1-d \leq m+1$ as claimed in item~(\ref{it1}).
We have
\[
f^{(m+1)}S-\varphi^{(m+1)}=
x^{-p}(f^{(m)}S-\varphi^{(m)})
-\mu(g^{(m)}S-\psi^{(m)}).\]
The degree of each term is $-d$ and the coefficients cancel.
Thus $\deg(f^{(m+1)}S - \varphi^{(m+1)}) < -d$.
We can see that item~(\ref{it2}) holds since  $-(m+1) + \deg ( f^{(m+1)} ) -1
= -d-1$.
The algorithm sets $g^{(m+1)}= \mu^{-1}f^{(m)}$ and
$\psi^{(m+1)}=\mu^{-1}\varphi^{(m)}$.
Item~(\ref{it4}) holds since
$(m+1)- \deg (f^{(m+1)} ) = d = \deg(g^{(m)})$.
Item~(\ref{it3}) holds since
$g^{(m+1)}S-\psi^{(m+1)}=\mu^{-1}(f^{(m)}S-\varphi^{(m)})$, which
has degree exactly $d-m-1 = -\deg(f^{(m+1)})$ and it is monic.
\end{proof}

The next few results show that the algorithm produces the minimal solution to the key equation, 
$f^e$ and $f^eS$.

\begin{lemma}
For all $m$, $\deg{f^{(m)}}\leq t$.
\end{lemma}

\begin{proof}
Consider $f^eg^{(m)} S-f^e\psi^{(m)}$.  This is a polynomial since
$f^eS$, $g^{(m)}$, $\psi^{(m)}$ and $f^e$ are.  Since the degree of
$f^e$ is $t$, we have
$\deg{(f^eg^{(m)}S-f^e\psi^{(m)})}= t-\deg{f^{(m)}}$,
using item~(\ref{it3}) in Theorem~\ref{t:BM}.  Thus $t-\deg{f^{(m)}}
\geq 0$.
\end{proof}

\begin{lemma}
\label{lemma:fmisfe}
When $m\geq 2t$,
$f^{(m)}=f^e$ and $\varphi^{(m)}=\varphi^{e}$.
\end{lemma}

\begin{proof}
Theorem~\ref{t:BM} tells us that $f^{(m)}S$ has no terms of degree
$-1,\dots,-m+\deg(f^{(m)})$.
From the previous lemma, if $m\geq 2t$ then $-m+\deg(f^{(m)}) \leq  -2t+t=-t$.
Thus, $f^{(m)} S$ has no terms of degree
$-1,\dots,-t$.
By Proposition~\ref{p:locator characterization},
$f^{(m)}$ must be a multiple of $f^e$;
by Theorem~\ref{t:BM} it is monic; and, by the preceding lemma, 
its degree is at most $t$. 
Thus, it must be equal to $f^e$.

On the other hand,
$\deg(f^eS-\varphi^{(m)})\leq -m+t-1\leq -t-1<0$.
Since both $f^eS$ and $\varphi^{(m)}$ are polynomials,
this means $\varphi^{(m)}=f^eS=\varphi^e$.
\end{proof}

\begin{proposition}
\label{p:BMworks}
If $t\leq\frac{d-1}{2}$ then the previous algorithm
outputs $f^e$ and~$\varphi^e$.
\end{proposition}

\begin{proof}
If $t\leq\frac{d-1}{2}$ then $n-k\geq d-1\geq 2t$ and the result follows from
Lemma~\ref{lemma:fmisfe}.
\end{proof}

\subsection{Error evaluation without the evaluator polynomial}
\label{s:RSeval}
We now derive a formula for error evaluation that does not use the
error evaluator polynomial, and thereby removes the need for computing
it.
It is called the Horiguchi-K\"otter algorithm in \cite{Blahut:Book} and
appears in \cite{horiguchi,kot97al}.

From the algorithm it is clear that
$$\left(\begin{array}{cc}
f^{(m)}&\varphi^{(m)}\\g^{(m)}&\psi^{(m)}\\
\end{array}\right) = 
U^{(m-1)} U^{(m-2)} \dots U^{(1)} U^{(0)}\left(\begin{array}{cc}
1 & 0\\0&-1\\
\end{array}\right)$$

Taking determinants, since each $U^{(m)}$ has determinant $1$, we get
\begin{align}
\label{e:RSdet}
f^{(m)}\psi^{(m)}-g^{(m)}\varphi^{(m)}&=-1.
\end{align}
In particular, when $m\geq 2t$,
$$f^{e}\psi^{(m)}-g^{(m)}\varphi^{e}=-1.$$
Let $j$ be such that $e_j\neq 0$. Evaluating at $\alpha_j$ we get
$g^{(m)}(\alpha_j)\varphi^{e}(\alpha_j)=1$ and so
$\varphi^{e}(\alpha_j)=(g^{(m)}(\alpha_j))^{-1}$.
Using Lemma~\ref{lemma:errorvalues}
we can establish the following proposition.

\begin{proposition}
\label{p:RSeval}
For $m\geq 2t$ and $g^{(m)}$ as in the Berlekamp-Massey
algorithm, if $e_j\neq 0$ then
$$e_j=({f^e}'(\alpha_j)g^{(m)}(\alpha_j))^{-1}.$$
\end{proposition}

The last proposition tells us that in the Berlekamp-Massey
algorithm we do not need to multiply $U^{(m)}$
by all the matrix $$\left(\begin{array}{cc}
f^{(m)}&\varphi^{(m)}\\g^{(m)}&\psi^{(m)}\\
\end{array}\right)$$
but by the vector
$$\left(\begin{array}{c}
f^{(m)}\\g^{(m)}\\
\end{array}\right).$$
Then the initialization step will be
$$\left(\begin{array}{c}
f^{(0)}\\g^{(0)}\\
\end{array}\right)
=
\left(\begin{array}{c}
1\\0\\
\end{array}\right)$$
and the updating step will be
$$\left(\begin{array}{c}
f^{(m+1)}\\g^{(m+1)}\\
\end{array}\right)
=U^{(m)}
\left(\begin{array}{c}
f^{(m)}\\g^{(m)}\\
\end{array}\right).$$

\subsection{Connections to the Euclidean algorithm}

Suppose that all the $\al_i$ defining the GRS code are $n$th roots of
unity.  In particular, we could demand that  none of the $\al_i$ are
zero and take $n=q-1$.
From the definition of $S$ it is easy to see that $s_a= s_{n+a}$ for
all $a \geq 0$, and consequently, 
\begin{align*}
S(x^n-1) &= s_0x^{n-1}+ s_1 x^{n-2} +
\cdots+s_{n-2}x + s_{n-1}.
\end{align*}
Call this polynomial $\barS$.
We might alter our definition of the key equation to say
that $f$, and $\varphi$ are solutions  when 
$f \barS = \varphi (x^n-1)$.  
That is
\begin{align}\label{e:euclid1}
f(s_0x^{n-1}+ s_1 x^{n-2} + \cdots+s_{n-2}x + s_{n-1} )
= \varphi( x^n-1).
\end{align}
Of course, the solution set is  the same as for our original equation,
and $f^e$, $\varphi^e$ are the minimal degree solutions such that $f^e$ is monic.
The result analogous to Theorem~\ref{t:BM} states that 
$\deg{(f^{(m)}\bar{S}-\varphi^{(m)})}\leq n-1-m+\deg{f^{(m)}}$ and
$g^{(m)}S-\psi^{(m)}$ is monic of degree
$n-\deg{f^{(m)}}$.
When the weight of $e$ is $t$, 
$f^{(2t)} = f^e$ and $\varphi^{(2t)}= \varphi^e$ give the least common
multiple of $\barS$ and $x^n-1$; the lcm is $f^eS = \varphi^e(x^n-1)$.

For a linear algebra perspective, write 
$f = f_0 + f_1x+ \dots +f_t x^t$.  The key equation requires that 
$\sum_{i=0}^t f_i s_{i+a} = 0$ for all $0 \leq a \leq n-t-1$.  
Setting $f_t=1$, there are $t $ unknowns, $f_0, \dots,f_{t-1}$, so 
the $t$ equations where $a=0,\dots,t-1$, are enough 
to determine the coefficients of $f$.  Thus we need to know the
syndromes $s_0$ to $s_{2t-1}$ to compute $f^e$. 
This verifies that the Berlekamp-Massey algorithm used with 
a code of redundancy  $2t$ can correct $t$ errors.

Equation~\eqref{e:euclid1} leads to a relationship between the
Berlekamp-Massey algorithm and the Euclidean algorithm.  
Let $m_0, m_1, \dots , m_r$ be the iterations of the algorithm in
which $p^{(m)}<0$
and let $m_{r+1}= 2t$ where $t$ is the weight of $e$.  
One can check that $p^{(m)}< - p^{(m_\ell)}$ for all
$m_\ell< m \leq m_{\ell+1}$.
For each $\ell=0,\dots,r$, let 
\begin{align*}
V^{(\ell)} &= U^{(m_{\ell+1}-1)}\cdots
U^{(m_\ell+1)}U^{(m_\ell)}. \\
\intertext{Then} 
V^{(\ell)} &= \begin{pmatrix}
q_{\ell} & -\mu^{(m_{\ell})} \\
( \mu^{(m_{\ell})})^{-1} &0 \\
\end{pmatrix}
\end{align*}
where $q_{\ell}$ is a monic polynomial of degree $p^{(m_{\ell})}$.
Define recursively, 
\begin{align*}\begin{pmatrix}
A_{0} \\
B_{0}
\end{pmatrix}
&= \begin{pmatrix}
\barS \\
x^n-1
\end{pmatrix} \qquad \text{ and} \\
\begin{pmatrix}
A_{\ell+1} \\
B_{\ell+1}
\end{pmatrix}
&= V^{(\ell)} \begin{pmatrix}
A_{\ell} \\
B_{\ell}
\end{pmatrix}
\end{align*}
so that $A_{\ell+1} = -\mu^{(m_\ell)} B_{\ell} + q_\ell A_{\ell}$
and
$B_{\ell +1} = (\mu^{(m_\ell)})^{-1}  A_\ell$.
Rearranging, we
get 
  $B_{\ell+1}=\frac{-\mu^{(m_{\ell-1})}}{\mu^{(m_\ell)}}\left(B_{\ell-1}-q_{\ell-1}B_{\ell}\right)$.  
This is a variant of the classical Euclidean algorithm for computing
the greatest common divisor with the modification that the remainders
are all monic.
We will sketch the main points and leave verification of the details
to the reader.

Notice that 
$A_\ell= f^{ (m_{\ell}-1) } \barS -\varphi^{( m_ \ell -1 )}(x^n-1)$
and similarly
$B_\ell= g^{ (m_\ell -1) }\barS -\psi^{(m_{\ell} -1 )}(x^n-1)$.
From the discussion after \eqref{e:euclid1},
$\deg B_\ell = n - \deg f^{(m_\ell -1)}$.  Referring to the
Berlekamp-Massey algorithm, $\deg f^{(m_{l+1}-1)} =\deg f^{(m_\ell)} 
< \deg f^{(m_{\ell}-1)}$ so we have $\deg B_{\ell+1} < \deg B_\ell$ and
the sequence of $B_\ell$ does indeed satisfy the requirements of the
Euclidean algorithm with monic quotients.  

At the final iteration, $m_{r+1}=2t$,
$f^{(m_{r+1})}= f^{e}$ and $\varphi^{(m_{r+1})} =
\varphi^e$ so that $A_{r+1}= f^e\barS - \varphi^e(x^n-1)=0$. 
As noted earlier, $f^e\barS$ is a constant multiple of the  lcm of
$\barS$ and $x^n-1$.
We also have $B_{r+1}= g^{(2t)}\barS - \psi^{(2t)}(x^n-1)$ is the
monic greatest common divisor of $\barS$ and $(x^n-1)$, namely 
$\prod_{i:e_i =0} (x-\al_i)$.

Thus we see that the Berlekamp-Massey algorithm breaks each division
of this version of the Euclidean algorithm  into several
steps, one for each subtraction of a  monomial multiple of the
divisor.  The Berlekamp-Massey algorithm is also more efficient than
the Euclidean algorithm, because it never computes the $B_\ell$.
It takes advantage of the fact that $B_0=x^n-1$ is very sparse, and
just computes the critical coefficients 
$\mu^{(m)}$ via the  polynomials $f^{(m)}$ and $\barS$. 

Berlekamp's formulation of the key equation was different from the one
presented here.  To obtain his formulation, let
\begin{align*}
\sig^e &= x^t f^e\left(\frac{1}{x}\right) = \prod_{k: e_k \not=0}(1-\al_k x) 
\\
\om^e &= x^{t-1} \varphi \left(\dfrac{1}{x}\right) 
=\sum_{j:e_j\neq 0}e_j\prod_{\substack{k:e_k\neq 0 \\k\neq j
  }}(1-\alpha_k x).
\end{align*}
These polynomials are $\Lambda(x)$ and $\Gam(x)$ respectively in
\cite{Blahut:Book,Roth:Book}.
Then \begin{align*}
x^{n+t-1}  f^e\left(\dfrac{1}{x}\right) \barS \left(\dfrac{1}{x}\right) 
&=  x^{n+t-1}\varphi^e\left(\dfrac{1}{x} \right) \left( \left(\dfrac{1}{x}\right)^n -1\right) \\
\sig^e \left(s_0+s_1x + \cdots + s_{n-1} x^{n-1}\right) &= \om^e (1-x^n) \\
\sig^e \left(s_0+s_1x + \cdots + s_{2t-1} x^{2t-1}\right) &\equiv \om^e \mod x^{2t}
\end{align*}
This is essentially the key equation
in~\cite{Berlekamp:Book,Blahut:Book,Roth:Book}, modulo  minor
changes due to different choices of  parity check matrix.

The  algorithm of Sugiyama et al~\cite{sugiyama:key} is based on  the equation  
\[\sig^e \left(s_0+s_1x + \cdots + s_{2t-1} x^{2t-1}\right)   + x^{2t}T = \om^e \]
One can run the Euclidean algorithm on 
$R_0=x^{2t}$ and $R_1=s_0 + \dots + s_{2t-1}x^{2t-1}$
until the remainder has degree less than $t$.
Sugiyama et al showed that the resulting combination of 
$(s_0+s_1x + \cdots + s_{2t-1} x^{2t-1})$ and $x^{2t}$ obtained is
$\om^e$ and that the coefficient of $(s_0+s_1x + \cdots + s_{2t-1}
x^{2t-1})$  is $\sig^e$.
The article~\cite{sugiyama:key} actually treats the more general situation of Goppa
codes and error-erasure decoding.

\section{The key equation for Hermitian codes}

The most widely studied algebraic geometry codes are those from
Hermitian curves.  One reason for the interest in Hermitian curves is that they are
maximal curves, meeting the
Weil bound on the number of points for a given genus.
They also have a very simple formula, and a great deal of symmetry,
which leads to lots of
structure that makes them useful in coding.
The short articles of Stichtenoth \cite{Stich:Herm} and Tiersma
\cite{Tiersma}, and Stichtenoth's book \cite{Stich:Book} are good
references for information on Hermitian curves and codes.

In this section we will derive the key equation and the algorithm for
solving it in a manner that parallels the section on Reed-Solomon
codes.  We will not discuss one very important issue:  The syndromes computed from
the received vector are insufficient for exploiting the full error correction capability of the
Berlekamp-Massey-Sakata decoding algorithm.  The majority voting algorithm of
Feng-Rao~\cite{FengRao} and Duursma~\cite{Duursma}  is required to compute
more syndrome values.
We will not discuss majority voting.
Instead, we simply deal with the problem solved by the BMS algorithm, computing
the error locator ideal from the syndrome of the error vector.
A detailed treatment of majority voting may be found in the chapter on
algebraic geometry codes by H{\o}holdt et al \cite{Hoholdt:Chap}.
The conditions ensuring success in the  majority voting algorithm  are
best understood in terms of the 
``footprint'' of the error vector, which is discussed below, and can lead to decoding 
beyond the minimum distance \cite{BrasOS:Correct}.

\subsection{The Hermitian curve}

Let $q$ be a prime power.  We will use the following equation for the
Hermitian curve over $\F_{q^2}$,
\[X^{q+1}=Y^q+Y.\]
For each $\alpha\in{\mathbb F}_{q^2}$, $\alpha^{q+1}$ is
the norm of $\alpha$ with respect to the extension ${\mathbb F}_{q^2}/{\mathbb F}_q$,
so  $\alpha^{q+1}$ belongs to ${\mathbb F}_q$.
On the other hand, $\beta^q+\beta$ is
the trace of $\beta$ with respect to $\F_{q^2}/\F_q$, so $\beta^q+\beta$ also belongs to $\F_q$.
Each element $\gam \in \F_q$ has $q$ preimages under the trace map, and $\gam$ has $q+1$  preimages
under the norm map (unless $\gam=0$ when there is one).  Thus 
there are  $n=q + (q-1)q(q+1) = q^3$ points on the curve. We label them 
$P_1=(\alpha_1,\beta_1),P_2=(\alpha_2,\beta_2),\dots,P_n=(\alpha_n,\beta_n)$.

Let ${\mathbb F}_{q^2}[X,Y]/(X^{q+1}-Y^q-Y)={\mathbb F}_{q^2}[x,y]$, where
$x$ is the image of $X$ in the quotient and $y$ is the image of $Y$.
Since $y^{q}=x^{q+1}-y$, each element $f$ in ${\mathbb F}_{q^2}[x,y]$
can be expressed in a unique way as a sum
$f_0(x)+f_1(x)y+f_2(x)y^2+\dots+f_{q-1}(x)y^{q-1}$. That is,
$\{1,y,\dots,y^{q-1}\}$ is a basis of ${\mathbb F}_{q^2}[x,y]$ as an
${\mathbb F}_{q^2}[x]$-module.
Also, ${\mathcal M}=\{x^ay^b:0\leq a,0\leq b<q\}$
is a basis of ${\mathbb F}_{q^2}[x,y]$ as a ${\mathbb F}_{q^2}$-vector space.

We wish to introduce a function on $\F_{q^2}[x,y]$ akin to the
degree function on $\F_q[x]$.  
Notice that any weighted degree in ${\mathbb F}_{q^2}[X,Y]$
such that $X^{q+1}$ and $Y^{q}+Y$ have equal weights
is obtained by assigning to $X$ a weight $kq$ and to $Y$ a weight $k(q+1)$
for some non-negative integer $k$.
Letting $k=1$, we define the {\it order function} $\rho$ by 
$\rho(x^ay^b)=\deg_{(q,q+1)}(X^aY^b)=aq+b(q+1)$ and
for $f=\sum_{a,b\geq 0}f_{a,b}x^ay^b$ we define
$\rho(f)=\max_{f_{a,b}\neq 0}\rho(x^ay^b)$.

One can see that $x^ay^b$ and $x^{a'}y^{a'}$ in ${\mathcal M}$ satisfy
$\rho(x^ay^b)=\rho(x^{a'}y^{a'})$ if and only if $a=a'$ and $b=b'$
and that $\rho({\mathbb F}_{q^2}[x,y])=\rho({\mathcal M})$.
Define $\Lambda=\rho({\mathbb F}_{q^2}[x,y])=
q{\mathbb N}_0+(q+1){\mathbb N}_0$.
The map $\rho:{\mathbb F}_{q^2}[x,y]\rightarrow\Lambda$ satisfies
$\rho(fg)=\rho(f)+\rho(g)$.
This suggests extending it to the quotient field of 
${\mathbb   F}_{q^2}[x,y]$, which we will write 
${\mathbb F}_{q^2}(x,y)$,
by defining $\rho(f/g)=\rho(f)-\rho(g)$.
Now the image of $\rho$ is all of ${\mathbb Z}$.

\subsection{Hermitian codes}

We define the evaluation map 
\begin{align*}
\ev:\F_{q^2}[x,y] &\longrightarrow \F^n \\
f &\longmapsto (f(\alpha_1, \beta_1),f(\alpha_2,\beta_2),\dots,
f(\alpha_n,\beta_n) ).
\end{align*}
The Hermitian code $H(m)$ 
over ${\mathbb F}_{q^2}$
is the linear code generated by $\{(f(P_1),\dots,f(P_n)):f\in{\mathcal M},\rho(f)\leq m\}$.
It is shown in \cite{Stich:Herm} (see also   \cite{JustesenHoholdt:book})
that $H(m)=\F_{q^2}^n$ when $m \geq q^3+q^2-q-1$ and that for $m<q^3+q^2-q-1$
the dual of $H(m)$ is $H(q^3+q^2-q-2-m)$.
Clearly, the monomials $x^ay^b$ such that $0 \leq b < q$ and $aq+b(q+1) \leq m$
are a basis for the space $\{f\in \F_{q^2}: \rho(f)\leq m\}$, so they may be used to create a generating
matrix for $H(m)$.
Since  $x^{q^2}-x$ vanishes on all points $P_k$, we should not use monomials $x^ay^b$ with
$a\geq q^2$ in the generating matrix.
This is only an issue when $m\geq q^3$.
Thus  for $m\in\Lambda$ and $m=aq+b(q+1)$, with $b<q$,
a generator matrix of $H(m)$ is obtained by evaluating monomials $x^{a'}y^{b'}$ whose
weighted degree is at most $m$ and such that $a' < q^2$.
$$\left(\begin{array}
{cccccc}
1 & 1 &&\dots&& 1\\
\alpha_1 & \alpha_2 && \dots && \alpha_{n}\\
\beta_1 & \beta_2 && \dots && \beta_{n}\\
\alpha_1^2 & \alpha_2^2 && \dots && \alpha_{n}^2\\
\alpha_1\beta_1 & \alpha_2\beta_2 && \dots && \alpha_{n}\beta_n\\
\beta_1^2 & \beta_2^2 && \dots && \beta_{n}^2\\
\vdots & \vdots && \vdots &&  \vdots \\
\alpha_1^a\beta_1^b &
\alpha_2^a\beta_2^b &&
 \dots &&
\alpha_n^a\beta_n^b \\
\end{array}\right).$$

\subsection{Polynomials for decoding}

Suppose that a word $c\in H(m)^\perp$ was transmitted and that a vector $u$ is received.
The vector  $e= u-c$ is the error vector.
Let $t$ be the weight of $e$.  
Define the {\em error locator ideal} of $e$ as
$$I^e=\{f\in{\mathbb F}_{q^2}[x,y]:f(\alpha_k,\beta_k)= 0\mbox{ for
  all } k\mbox{ with }e_k\neq0\}$$ 
and the {\em syndrome}  for $e$ as
\begin{align}
\label{e:Hsyn}
S=\sum_{k=1}^ne_k\frac{x^{q+1}-\alpha_k^{q+1}}{(x-\alpha_k)(y-\beta_k)}=
\sum_{k=1}^ne_k\frac{y^{q}+y-\beta_k^{q}-\beta_k}{(x-\alpha_k)(y-\beta_k)}.
\end{align}
Notice that the order
of each term in the summand is $q^2-q-1$.

We will give three justifications for this definition of the syndrome
in the lemmas below.
We note first that for any $(\alpha,\beta)\in{\mathbb F}_{q^2}^2$
on the Hermitian curve,
\begin{align*}
\frac{x^{q+1}-\alpha^{q+1}}{x-\alpha}&=
\alpha^{q}\frac{\left(\frac{x}{\alpha}\right)^{q+1}-1}{\frac{x}{\alpha}-1}\\
&=\alpha^q\left(\left(\frac{x}{\alpha}\right)^{q}+\left(\frac{x}{\alpha}\right)^{q-1}+\dots+\frac{x}{\alpha}+1\right)\\
&=x^q+\alpha x^{q-1}+\dots+
\alpha^{q-1}x+\alpha^q
\end{align*}
and
\begin{eqnarray*}
\frac{y^{q}+y-\beta^{q}-\beta}{y-\beta}&=&
1+\frac{y^{q}-\beta^{q}}{y-\beta}\\
&=&
1+y^{q-1}+\beta y^{q-2}+\dots+\beta^{q-2}y+\beta^{q-1}
\end{eqnarray*}
We will use these identities several times during this presentation.

The first lemma gives a nice relationship between $I^e$ and $S$.  We
will show later that the converse also holds.

\begin{lemma}
\label{l:fSinR}
If $f\in I^e$ then $fS\in {\mathbb F}_{q^2}[x,y]$.
\end{lemma}

\begin{proof}
If $f\in I^e$ and $P_{k_1},\dots,P_{k_t}$
are the error positions then
there exist $g_{k_1},\dots,g_{k_t}$ and $h_{k_1},\dots,h_{k_t}$ in ${\mathbb F}_{q^2}[x,y]$
such that
\begin{eqnarray*}
f&=&g_{k_1}(x-\alpha_{k_1})+h_{k_1}(y-\beta_{k_1})\\
&=&g_{k_2}(x-\alpha_{k_2})+h_{k_2}(y-\beta_{k_2})\\
&\vdots&\\
&=&g_{k_t}(x-\alpha_{k_t})+h_{k_t}(y-\beta_{k_t})\\
\end{eqnarray*}
Hence
\begin{eqnarray*}
fS&=&\sum_{k_i:e_{k_i}\neq 0}e_{k_i}
\left(
g_{k_i}\frac{y^{q}+y-\beta_{k_i}^{q}-\beta_{k_i}}{y-\beta_{k_i}}+h_{k_i}\frac{x^{q+1}-\alpha_{k_i}^{q+1}}{x-\alpha_{k_i}}
\right)\\
&=&\sum_{k_i:e_{k_i}\neq 0}e_{k_i}g_{k_i}
\left(1+y^{q-1}+\beta_{k_i}y^{q-2}+\dots+\beta_{k_i}^{q-2}y+\beta_{k_i}^{q-1}
\right)\\
&&+\sum_{k_i:e_{k_i}\neq 0}e_{k_i}h_{k_i}
\left(x^q+\alpha_{k_i}x^{q-1}+\dots+
\alpha_{k_i}^{q-1}x+\alpha_{k_i}^q \right)
\end{eqnarray*}
which belongs to ${\mathbb F}_{q^2}[x,y]$.
\end{proof}

The next lemma shows that for $f\in I^e$,
the product $fS$ may be used for error evaluation.
We will need the derivative of $y$ with respect to $x$.  Since $q=0$
in ${\mathbb F}_{q^2}$, and  $d(y^q+y)/dx=d(x^{q+1})/dx$, we deduce that
$dy/dx=x^q$.  We say that $f$ has a {\em simple zero} at a point $P$
when $f(P)=0$ but $f'(P) \not=0$.

\begin{lemma}
\label{l:Hevaluator}
If $f\in I^e$ and $P_k$ is an error position then  $e_kf'(P_k)=fS(P_k).$
If $f$ has a simple zero at  $P_k$ then
$$e_k=\frac{fS(P_k)}{f'(P_k)}.$$
\end{lemma}

\begin{proof}
The rational function $\frac{x^{q+1}-\alpha_j^{q+1}}{(x-\alpha_j)(y-\beta_j)}=
\frac{y^{q}+y-\beta_j^{q}-\beta_j}{(x-\alpha_j)(y-\beta_j)}$
gives a well defined value at any point different from
$(\alpha_j, \beta_j)$, so when $j\not=k$,
$\left(f\frac{x^{q+1}-\alpha_j^{q+1}}{(x-\alpha_j)(y-\beta_j)}\right)(P_k)=0$.
Consequently,
$$fS(P_k)=e_{k}\left(f\frac{x^{q+1}-\alpha_k^{q+1}}{(x-\alpha_k)(y-\beta_k)}\right)(P_k).$$
Since $f(P_k)=0$, there are $g,h\in{\mathbb F}_{q^2}[x,y]$ such that 
$f=(x-\alpha_k)g+(y-\beta_k)h$.
Hence,
\begin{eqnarray*}
fS(P_k)&=&
e_{k}\left((1+y^{q-1}+\beta_{k}y^{q-2}+\dots+\beta_{k}^{q-2}y+\beta_{k}^{q-1})(P_k)\right)g(P_k).
\\&& +e_k \left((x^q+\alpha_{k}x^{q-1}+\dots+
\alpha_{k}^{q-1}x+\alpha_{k}^q)(P_k)\right)h(P_k)\\
&=&e_k\left((1+q\beta_k^{q-1})g(P_k)+(q+1)\alpha_k^qh(P_k)\right)
\\
&=&e_k\left(g(P_k)+\alpha_k^qh(P_k)\right).
\end{eqnarray*}
On the other hand,
$f'=
g+x^qh+(x-\alpha_k)g'+(y-\beta_k)h'$.
Evaluating $f'$ at $P_k$, 
\begin{align*}f'(P_k)&=g(P_k)+\alpha_k^qh(P_k), \text{ so } \\
e_k f'(P_k) &= fS(P_k). \\
\intertext{When $f$ has a simple zero at $P_k$}
e_k&=fS(P_k)/f'(P_k).
\end{align*}
\end{proof}

As our final justification for our definition of $S$, we show that the
syndrome values for the vector $e$, that is the products $\ev(x^ay^b) \cdot e$,
appear as coefficients in a particular expansion of $S$.

\begin{lemma}
 Let $s_{a,b}= \sum_{k=1}^n e_k \alpha_k^a \beta_k^b$ and
let $\delta_b$ be 1 when $b=0$ and $0$ otherwise.
\[ S=  \frac{1}{x}\sum_{b=0}^{q-1} \sum_{a=0}^\infty  s_{a,b} x^{-a}(y^{q-1-b}+\delta_b)\]
\end{lemma}

\begin{proof}

\begin{eqnarray*}
\frac{y^{q}+y-\beta_k^{q}-\beta_k}{(x-\alpha_k)(y-\beta_k)}&=&
\left(1+\frac{y^q-\beta_k^q}{y-\beta_k}\right)\frac{1}{x}\left(\frac{1}{1-\frac{\alpha_k}{x}}\right)\\
&=&(1+y^{q-1}+\beta_ky^{q-2}+\dots+\beta_k^{q-2}y+\beta_k^{q-1})(\frac{1}{x}+\frac{\alpha_k}{x^2}+\frac{\alpha_k^2}{x^3}+\dots)\\
&=&\sum_{0\leq a}\sum_{0\leq b< q}\alpha_k^a\beta_k^bx^{-a-1}(y^{q-1-b}+\delta_{b}).
\end{eqnarray*}

Hence,
\begin{eqnarray*}
S&=&\sum_{k=1}^ne_k\sum_{0\leq a}\sum_{0\leq b< q}\alpha_k^a\beta_k^bx^{-a-1}(y^{q-1-b}+\delta_{b})\\
&=&\sum_{0\leq a}\sum_{0\leq b<
  q}\left(\sum_{k=1}^ne_k\alpha_k^a\beta_k^b \right)x^{-a-1}(y^{q-1-b}+\delta_{b})\\
&=&\frac{1}{x}\sum_{0\leq a}\sum_{0\leq b< q}s_{a,b}x^{-a}(y^{q-1-b}+\delta_{b})\\
\end{eqnarray*}
\end{proof}

\subsection{Another basis for $\F_{q^2 }(x,y)$}
The final result of the previous section  suggests that we introduce a new
basis for  $\F_{q^2}(x,y)$ over $\F_{q^2}(x)$. 
For $0 \leq b < q$, let
\begin{align}
\label{eq:Hdual} 
\zstar_b &= \begin{cases}  y^{q-1} +1  & \text{ if } b= 0 \\
                         y^{q-1-b}  & \text{ otherwise. }
            \end{cases}
            \end{align}
Notice that $\rho(\zstar_b)= (q+1)(q-1-b) = q^2-1-b(q+1)$.
We will call $\{\zstar_b: b = 0,\dots,q-1\}$ the $\ast$-basis.
We will write the syndrome, and products 
of the syndrome with polynomials in $x,y$, using the $\ast$-basis.
An element $f\in \F_{q^2}[x,y]$ is {\em monic} in the $\ast$-basis
when its leading term, say $f_{a,b}x^a\zstar_b$, has $f_{a,b}=1$.
The following two  lemmas show how this basis is useful for decoding.

\begin{lemma}
\label{l:Hdual}
The coefficient of $\zstar_0$ in $y^b\zstar_c$ is $1$ if
$b=c$ and is 0 otherwise.
\end{lemma}

\begin{proof}
One can prove by a straightforward computation that
for $0 \leq b,c < q$, 
\begin{align*}
y^b\zstar_c &=
             \begin{cases}  \zstar_0  & \text{ if } b=c= 0 \\
                          \zstar_0 -\zstar_{q-1} & \text{ if } b=c\not=0 \\
                          x^{q+1}\zstar_{q-b} & \text{ if } b>c=0 \\
                          x^{q+1}\zstar_{q+c-b} - \zstar_{q-1+c-b}& \text{ if } b>c>0 \\
                          \zstar_{c-b} & \text{ if } c>b \\
            \end{cases}
\end{align*}

For example, if $b>c>0$ then
\begin{alignat*}{2}
y^b \zstar_c &= y^{q-1+b-c}\\
  & = y^{b-c-1}(x^{q+1}-y)  \\
  &= x^{q+1}\zstar_{q+c-b} - \zstar_{q-1+c-b}
\end{alignat*}
Notice that $2\leq q+c-b \leq q-1$, so that each of the indices in
this case is between $1$ and $q-1$.   Thus the coefficient of $\zstar_0$ in 
$y^b\zstar_c$ is 0 when $0 \leq b\leq c$. 
Similar arguments apply to the other cases.
\end{proof}

Any element of $\F_{q^2}(x,y)$ may be  expressed uniquely as
$\sum_{b=0}^{q-1} h_b \zstar_b $ for $h_b \in \F_{q^2}(x)$.
We will write $h_b$ in the form used for the syndrome in
Section~\ref{s:RS}, 
$h_b= \frac{1}{x} \sum_a h_{a,b}x^{-a}$,  where it is
understood that $a$ varies over 
all integers larger than some unspecified bound.
For example, we will write the syndrome as 
$S=\frac{1}{x} \sum_{b=0}^{q-1}\sum_a s_{a,b} x^{-a}\zstar_b$,
where  it is understood that $s_{a,b}=0$ for $a<0$.

\begin{lemma}
\label{l:t_ab}
Let $f \in \F_{q^2}[x,y]$, let $a \in \Z$ and let $b$ satisfy $0 \leq b <q$.
The coefficient of  $x^{-a-1}\zstar_b$ in $fS$
equals the coefficient of $\zstar_0/x$ in $x^ay^bfS$.

More precisely, expand $\tf= y^bf$,  $S$, and $fS$ as follows.
\begin{align*}
\tf  &= \quad\sum_{c=0}^{q-1} \tf_c  y^c 
=\sum_{c=0}^{q-1} \sum_{a} \tf_{a,c}\, x^a y^c \\
S &= \frac{1}{x} \sum_{c=0}^{q-1} s_c \zstar_c
= \frac{1}{x} \sum_{c=0}^{q-1} \sum_{a} s_{a,c}\,x^{-a} \zstar_c \\
fS &= \frac{1}{x} \sum_{c=0}^{q-1} t_c \zstar_c
= \frac{1}{x} \sum_{c=0}^{q-1} \sum_a t_{a,c}\,  x^{-a} \zstar_c
\end{align*}
Here $s_c = \sum_a s_{a,c} x^{-a}$ and similar
definitions hold for $t_c$ and $\tf_c$.
Then
\begin{alignat*}{2}
t_b &= \sum_{c=0}^{q-1} \tf_c  s_c &\qquad \text{ and } \qquad
t_{a,b}&= \sum_{c=0}^{q-1} \sum_i \tf_{i,c}\, s_{i+a,c}. 
\end{alignat*}
\end{lemma}

\begin{proof}
From the previous lemma, the coefficient of $\zstar_0$ in
\begin{align*}
y^b(fS) &= \frac{1}{x} \sum_{c=0}^{q-1} t_c y^b \zstar_c \\
\intertext{is $(1/x) t_b$. On the other hand, $y^bf = \tf $, so}
 (y^bf)S &= \left(\sum_{c=0}^{q-1}\tf_c y^c\right)
\left(\frac{1}{x} \sum_{d=0}^{q-1} s_d \zstar_d \right)\\
&= \frac{1}{x} \sum_{c=0}^{q-1}\sum_{d=0}^{q-1} \tf_c s_d
y^c\zstar_d.
\end{align*}
Applying the previous lemma, the coefficient of $\zstar_0$ is $(1/x)
\sum_{c=0}^{q-1} \tf_c s_c$.
We conclude that $t_b = \sum_{c=0}^{q-1} \tf_c s_c$.
Writing $\tf_c = \sum_i \tf_{i,c}\, x^i$ and
$s_c = \sum_j s_{j,c}\, x^{-j}$
we have
\begin{align*}
\tf_cs_c &= \sum_{i} \tf_{i,c}\, x^i \sum_j s_{j,c}\, x^{-j} \\
&= \sum_a x^{-a} \sum_i \tf_{i,c}\, s_{i+a,c}.
\end{align*}
This sum is finite since $\tf_c$ has finite support, and it gives the
formula for $t_{a,b}$.
\end{proof}

This lemma tells us  that to identify the coefficient of
$x^{-a-1}\zstar_b$ in $fS$, we write  $\tf = y^bf$ in the standard basis
and then compute the recursion
$t_{a,b}= \sum_{c=0}^{q-1} \sum_{i} \tf_{i,c}\, s_{i+a,c} $.

\subsection{The key equation}

We now define the key equation and  approximate solutions to the key
equation.  We establish some simple lemmas that show basic
properties of approximate solutions and how two approximate
solutions can be combined to get a better approximation.

\begin{definition}
\label{d:Hkey}
We say that $f, \varphi \in \F_{q^2}[x,y]$ solve the key equation for
syndrome $S$ when $fS=\varphi$.
\end{definition}

For a nonzero $f \in \F_{q^2}[x,y]$, writing $fS = 
\frac{1}{x} \sum_{b=0}^{q-1} \sum_a t_{a,b}\, x^{-a}\zstar_b$, 
we see that $f, \varphi$ satisfy the key equation
when $t_{a,b}=0$ for $a\geq 0$ and $\varphi = \frac{1}{x}
\sum_{b=0}^{q-1} \sum_{a<0} t_{a,b}\, x^{-a}\zstar_b$.

 \begin{definition}
\label{d:Happroxkey}
We say that $f$ and $ \varphi$ in $\F_{q^2}[x,y]$, with $f$ nonzero,
solve the $K$th~ approximation  
of the key equation for syndrome $S$ (or the $K$th key equation, for
short) 
when the following two conditions hold.
\begin{enumerate}
\item $\rho( fS-\varphi) \leq q^2-q-1-K$,
\item 
$\varphi$,  written in the $\ast$-basis, is a sum of terms whose
order is at least $q^2-q-K$.
\end{enumerate}
We will also say that $0$ and $x^{-a-1}\zstar_b$, for $a<0$,
solve the $aq+b(q+1)$ key equation.  
\end{definition}

Notice that  $ \rho(x^{-a-1}\zstar_b) = q^2-q-1-(aq+b(q+1))$, 
and when $a<0$, we have  $x^{-a-1}\zstar_b \in\F_{q^2}[x,y]$.
Thus, for $0, x^{-a-1}\zstar_b$, condition (1) holds with $K=aq+b(q+1)$,
but condition~(2) is not satisfied.   It is convenient to make
this pair a solution to the key equation, so we have included the
special case in the definition.

For $f \not=0$, 
(1) means that $fS-\varphi$ has only terms $x^{-a-1}\zstar_b$ with
$aq+b(q+1)\geq K$, while 
(2) means that $\varphi$ has only terms 
$x^{- a-1}\zstar_b$ with $aq+b(q+1)< K$ and with $a<0$ because
$\varphi$ is a polynomial.  
Consequently, 
using the expression for $fS$ above, 
$f, \varphi$ solve the $K$th key equation
if and only if
\begin{align*}
&t_{a,b}=0 \text{ whenever }a\geq 0 \text{  and } aq+b(q+1) < K,
\text{ and } \\
&\varphi = \frac{1}{x} \sum_{b=0}^{q-1} 
\sum_{\substack{a <0\\ aq+b(q+1)< K}} t_{a,b}x^{-a}\zstar_b.
\end{align*}

\begin{example}
\label{x:Hbasestep}
The pair $y^b, 0$ satisfies the $-b(q+1)$ key equation.
 We have \[\rho(y^b S) \leq b(q+1) + q^2-q-1= q^2-q-1-(-b(q+1)).\]

The pair $0, \zstar_b$ satisfies the $b(q+1) -q$ key equation.
\[ \rho(\zstar_b) = (q-1-b)(q+1) = q^2-q-1-( b(q+1) -q) \]
\end{example}

The following technical lemmas will be used to simplify the proof 
of Theorem~\ref{t:HBMS}, which establishes the properties of the decoding algorithm.

\begin{lemma}
\label{l:gfS}
Suppose that $f \ne 0$ and that $f, \varphi$ satisfy the $K$th key
equation for syndrome $S$.
Let $g \in \F_{q^2}[x,y]$ with $\rho(g) <K$.
Then, in the $\star$-basis expansion of $gfS$ the coefficient of
$\zstar_0/x$ is $0$.  Consequently, if $g$ and $h$ are both monic of
order $K$ then the coefficients of $\zstar_0/x$ in $gfS$ and $hfS$ are equal.
\end{lemma}

\begin{proof}
It is sufficient to establish this result for a monomial, $g=x^ay^b$,
with $aq+b(q+1)  < K$.  
By Lemma~\ref{l:t_ab}, the coefficient of $\zstar_0/x$ in $x^ay^bfS$
is equal to the coefficient of $x^{-a-1}\zstar_b$ in $fS$.  
Expanding  $fS$ as in Lemma~\ref{l:t_ab}, this coefficient is
$t_{a,b}$.  The discussion after the
definition of approximate solutions to the key equation shows that
$t_{a,b}=0$ for  $a\geq 0$ and  $aq+b(q+1) <K$.  
The final statement of the lemma follows from $\rho(g-h) <K$.
\end{proof}

\begin{lemma}
\label{l:Hshift}
Suppose that $f, \varphi$ satisfy the $K$th key equation.  For any
nonnegative integer $i$, the   $K-iq$ key equation is satisfied by
$x^if,x^i\varphi$. 
\end{lemma}

\begin{proof}
It is trivial to check the lemma for the case when $f=0$ and $\varphi=
x^{-a-1}\zstar_b$.
For $f\not=0$, we certainly have $ x^i f, x^i\varphi \in \F_{q^2}[x,y]$.  
The terms in $\varphi$ have order at least $q^2-q-K$, so
the terms in $x^i\varphi$ have order at least $q^2-q-K +iq =
q^2-q-(K-iq)$. 
We also assume $\rho(fS-\varphi) \leq q^2-q-1-K$, so 
$\rho(x^i(fS-\varphi)) \leq q^2-q-1-(K-iq)$.
\end{proof}

Notice that an analogous result does {\em not} hold for multiplication
by $y$.  The example above shows that $0, \zstar_{1}$ solves the $K=1$ key
equation.  
Yet,  $0$, $y\zstar_1$ 
does not solve the  key equation of order $K-(q+1) = -q$.
Indeed, $y\zstar_1= \zstar_0 -\zstar_{q-1}$ and
the $-\zstar_{q-1}$ term violates the requirements of the definition.

\begin{lemma}
\label{l:Hupdate}
Suppose that $f, \varphi$ and $g, \psi$ satisfy the $K$th key equation
where $K= aq+b(q+1)$.  Suppose in addition that $f\not=0$ and
$gS-\psi$ is monic of order
 $q^2-q-1-K$.   Let the coefficient of
$x^{-a-1}\zstar_b$ in $fS$ be $\mu$.  Then $f-\mu g, \, \varphi-\mu
\psi$ satisfy the $(K+1)$th key equation. 
\end{lemma}

\begin{proof}
By assumption, $\rho(fS-\varphi) \leq q^2-q-1-K$ and $\varphi$ has terms of
order 
$q^2-q-K$ or larger.  Furthermore, $\mu$ is the coefficient of
$x^{-a-1}\zstar_b$ in $fS$.  Since $\rho(x^{-a-1}\zstar_b)=q^2-q-1-K$,
when $\mu=0$  the inequality above is
strict, and $f,\varphi$ solve the $(K+1)$th key equation.  
Suppose $\mu\ne 0$.  Then both  $fS-\varphi$ and  $gS-\psi$ have order 
$q^2-q-1-K$.  Since $gS-\psi$ is monic, 
$\rho((fS-\varphi)-\mu (gS-\psi)) < q^2-q-1-K$.  
Furthermore, $\psi$ has terms of order at least $q^2-q-1-K$ (allowing
for the case in which $g=0$) 
so $\varphi-\mu\psi$ has terms of order at  least $q^2-q-(K+1)$, as
required for the $(K+1)$th key equation.
\end{proof}

Proposition~\ref{p:Hlocator characterization} below is a
generalization of  Proposition~\ref{p:locator characterization} to
Hermitian curves.  It also gives
the converse of Lemma~\ref{l:fSinR}.  First, we need a lemma.

\begin{lemma}
\label{lemma:mustdivide}
Let $f\in{\mathbb F}_{q^2}[x,y]$ and let $(\alpha,\beta)\in{\mathbb
  F}_{q^2}$ be a point on the Hermitian curve. 
Then $f(\alpha,\beta)$ is the coefficient of $\zstar_0 /x$ in the 
$\ast$-basis expansion of $f\frac{x^{q+1}-\alpha^{q+1}}
{(x-\alpha)(y-\beta)}$.
\end{lemma}

\begin{proof}
We know that $f(x,y)=f(\alpha,\beta)+(x-\alpha)g+(y-\beta)h$ for some
$g,h\in{\mathbb F}_{q^2}[x,y]$.
Thus $f\frac{x^{q+1}-\alpha^{q+1}} {(x-\alpha)(y-\beta)}$
has a polynomial part plus
\begin{align}
f(\alpha,\beta)\frac{x^{q+1}-\alpha^{q+1}} {(x-\alpha)(y-\beta)} 
&=  \dfrac{f(\alpha,\beta)}{x-\alpha} \left( y^{q-1}+1 + \beta y^{q-2}
  + \dots + \beta^{q-2}y + \beta^{q-1} \right) \\
&= f(\alpha,\beta) \dfrac{1}{x}\sum_{b=0}^{q-1} \sum_a \alpha^a \beta^b
 x^{-a} \zstar_b
\end{align}
The coefficient of $\zstar_0/x$ is $f(\alpha, \beta)$ as claimed.
\end{proof}

We need some facts about generators for $I^e$.  
We summarize here material that is treated in depth in \cite{Hoholdt:Chap}.
Recall that $\Lambda = \{ \rho(f): f \in \F_{q^2}[x,y]\}$. 
Define the {\em footprint} of $e$ as $\Delta^e=\Lambda-\rho(I^e).$
The quotient ring ${\mathbb F}_{q^2}[x,y]/I^e$ 
is a $t$-dimensional ${\mathbb F}_{q^2}$-vector space.
A basis for this space is obtained by taking the classes
of $x^ay^b$
for $\rho(x^ay^b) \in \Delta^e$, so $|\Delta^e|=t$.  

Since $\F_{q^2}[x]$ is a principal ideal domain, for any ideal $I$
with $\F_{q^2}[x,y]/I$ finite dimensional over $\F_{q^2}$, the ideal
$I$ is a free module over $\F_{q^2}[x]$  of rank $q$.  
For each $i$ with $0 \leq i \leq q-1$
let $f_i $ be such that $\rho(f_i)$ is  minimal among 
$\{f \in I : \rho(f) \equiv i \mod q\}$.
Then $\{f_i : 0 \leq i \leq q-1\}$ is a Gr\"obner basis for $I$.
By reducing $f_i$ by multiples of $f_j$ for $j \ne i$ we may assume
that all nonzero terms of $f_i$, except the leading term, have
order in $\Delta= \Lam - \rho(I)$.

\begin{proposition}
\label{p:Hlocator characterization}
If the expansion of $fS$ in the $\ast$-basis has zero coefficients
for all $x^{-a-1}\zstar_b$ such that $aq+b(q+1) \in \Delta^e$ 
then $f\in I^e$.
In particular, let $K$ be the maximal element of $\Delta^e$.
If $f$, $\varphi$ satisfy the $(K+1)$th key equation then $f \in I^e$.
\end{proposition}

\begin{proof}
Let $f$ satisfy the hypotheses of the proposition.
Let $e_k\neq 0$ and let $P_k=(\alpha_k,\beta_k)$.
We will prove that $f(P_k)=0$.
Consider the ideal 
\[I' = \{ h\in \F_{q^2}[x,y] : h(P_j)= 0 \text{ for all } j \text{
  with } e_j \ne 0 \text{ and } j \ne k\}\]
and let $\Delta' = \Lambda\setminus \{\rho(f) : f \in I'\}$.
Notice that  $|\Delta'|=t-1$, so there exists some $g \in I'$ 
such
that $\rho(g) \in \Delta^e \setminus \Delta'$.
Reducing $g$ modulo  a reduced Gr\"obner basis for $I'$, 
we can  ensure that every monomial in $g$ has order in $\Delta^e$.

As in Lemma~\ref{l:t_ab}, write $fS = \frac{1}{x} \sum_{b=0}^{q-1}
\sum_a t_{a,b} x^{-a} \zstar_b$. 
Lemma~\ref{l:t_ab}
shows that $t_{a,b}$ is the coefficient of 
$\zstar_0/x$ in $x^ay^bfS$.  
For $aq+b(q+1) \in \Delta^e$, the hypothesis of this lemma is that
$t_{a,b} = 0$, so  the coefficient of  $\zstar_0/x$ in
$x^ay^bfS$ is~$0$.
Since $g$ is a linear combination of monomials with order in
$\Delta^e$, the coefficient of $\zstar_0/x$ in $gfS$ is~$0$.

On the other hand,  Lemma~\ref{lemma:mustdivide} and the definition of
the syndrome, \eqref{e:Hsyn}, shows that 
the coefficient of $\zstar_0/x$ in $gfS$ is 
\[ \sum_{j=1}^n e_j g(P_j)f(P_j) = e_k g(P_k)f(P_k). \]
Here we have used $g(P_j)= 0$ for $j\not= k$ since $g \in I'$.
Since  $g \not \in I^e$ we must have $g(P_k)\ne 0$, so we conclude
that $f(P_k)=0$.

The final statement of the proposition follows immediately from the
observation following the definition of approximate solutions to the
key equation.  If $f, \varphi$ satisfy the $K$th key equation for
$K=1+\max \Delta^e$, 
then the coefficient of $x^{-a-1}\zstar_b$ is 0 for any $a\geq 0$ and
$aq+b(q+1)< K$.
\end{proof}

\subsection{Solving the key equation}
As noted in the introduction, this algorithm is based on K\"otter's
version of Sakata's generalization of the Berlekamp-Massey algorithm.
The algorithm uses the algebra of $\F_{q^2}[x,y]$ in only one place,
the computation of $\tf$, otherwise all the computations involve
polynomials in $x$, which are easily implementable using shift-registers.

The value $M$ determining the final iteration of the algorithm is given in 
Proposition~\ref{p:HfinalM} below.

\begin{center}
{\bf Decoding algorithm for  Hermitian codes}
\end{center}

\noindent{\bf Initialize:}
For $i=0$ to $q-1$, set
$\left(\begin{array}{cc}
f_i^{(0)}&\varphi_i^{(0)}\\g_i^{(0)}&\psi_i^{(0)}\\
\end{array}\right)=
\left(\begin{array}{cc}y^i&0\\0& -\zstar_i\\
\end{array}\right)$

\medskip

\noindent {\bf Algorithm:} For $m=0$ to $M$, and for each pair $i,j$ such that $m\equiv i+j\mod q$, set

\begin{tabular}{ll}
$d_i=\rho({f_i^{(m)}})$ & $  \qquad d_j = \rho({f_j^{(m)}})$\\ \medskip

$r_i=\frac{m-d_i - j(q+1)}{q} $&$ \qquad r_j=\frac{m-d_j - i(q+1)}{q} $\\ \medskip

$\tf_i =y^j f_i $&$ \qquad \tf_j = y^if_j$\\  \medskip

$\mu_i = \sum_{c=0}^{q-1} \sum_a (\tf_i)_{a,c} s_{a +r_i,c} $&$ \qquad 
\mu_j = \sum_{c=0}^{q-1} \sum_a (\tf_j)_{a,c} s_{a+r_j,c}$\\  \medskip

$p=\frac{d_i+d_j -m }{q} -1$ &
\end{tabular}

\medskip
The update for $j$ is analogous to the one for $i$ given below.\\

$U_{i}^{(m)}=\left\{\begin{array}{ll}
\left(\begin{array}{cc}
1 & -\mu_i x^p\\
0 & 1\\
\end{array}\right)& \mbox{ if }\mu_i= 0 \mbox{ or }p\geq 0
\\
\left(\begin{array}{cc}
x^{-p}&-\mu_i\\
1/\mu_i& 0
\end{array}\right)
& \mbox{ otherwise.}
\\
\end{array}
\right.$

\medskip
\begin{quote}
$\left(\begin{array}{cc}
f_i^{(m+1)}&\varphi_i^{(m+1)}\\g_j^{(m+1)}&\psi_j^{(m+1)}\\
\end{array}\right)
=U_i^{(m)}\left(\begin{array}{cc}
f_i^{(m)}&\varphi_i^{(m)}\\g_j^{(m)}&\psi_j^{(m)}\\
\end{array}\right)$
\end{quote}

\noindent {\bf Output:} $f_i^{(M+1)}, \varphi_i^{(M+1)}$ for $0 \leq i <q$.
 
\medskip

\begin{remark}
The monomial $x^{r_i}y^j$ used to define $\widetilde{f}_i$
is the {\em shift} necessary so that $x^{r_i}y^jf_i^{(m)}$ has leading term
of order $m$. Indeed,
$\rho(x^{r_i}y^jf_i^{(m)})=r_iq+j(q+1)+d_i=m.$
Lemma~\ref{l:t_ab} says that $\mu_i$ is the coefficient of
$x^{-r_i-1}\zstar_j$ in $f_i^{(m)}S$. 
\end{remark}

\begin{theorem}
\label{t:HBMS}
For $m\geq 0$, 
\begin{enumerate}
\item\label{ith1}
$f_i^{(m)}$ is monic and $\rho(f_i^{(m)})\equiv i \mod q$.
\item\label{ith2}
$f_i^{(m)}, \varphi_i^{(m)}$ satisfy the $m-\rho(f_i^{(m)})$ approximation of the key equation. 
\item\label{ith3}
$g_i^{(m)}, \psi_i^{(m)} $  satisfy the $\rho(f_i^{(m)}) -q$
approximation of the key equation and $g_i^{(m)}S-\psi_i^{(m)}$ is monic
of order
$q^2-1-\rho(f_i^{(m)})$.
\item\label{ith4}
$\rho{(g_i^{(m)})} <  m-\rho{(f_i^{(m)})}+q$.
\end{enumerate}
\end{theorem}

\begin{proof}
We will proceed by induction on $m$.
Example~\ref{x:Hbasestep} establishes the base step, $m=0$.

Assume that the statements of the theorem are true for $m$, we
will prove them for $m+1$.  It is sufficient to consider a pair $i,j$
with $0 \leq i,j < q-1$  satisfying $i+j \equiv m \mod q$.  
Let $d_i$, $r_i$, $\mu_i$, $p$ be as defined in the algorithm.

The induction hypothesis says that $f_i^{(m)}, \varphi_i^{(m)}$ satisfy the
$m-d_i$ key equation.
By   Lemma~\ref{l:t_ab}, $ \mu_i$ is the coefficient
of $x^{-r_i-1}\zstar_j$ in $f_i^{(m)}S$. 
A simple computation shows  
$\rho(x^{-r_i-1}\zstar_j)= q^2-q-1-(m-d_i)$.
Consequently, if $\mu_i=0$, then  $f_i^{m}, \varphi_i^{m}$ solve the $(m+1-d_i)$ key
equation. 
In this case, the algorithm retains the data from the iteration $m$, 
{\em e.g.} $f^{(m+1)} = f^{(m)}$.
It is easy to verify that the properties of the  theorem hold.

If $\mu_i\not=0$, we consider two cases.  First, suppose  $p\geq 0$.
The algorithm sets $f_i^{(m+1)}=f_i^{(m)}-\mu_ix^pg_i^{(m)}$.
Notice that    $\rho{(\mu_ix^pg_i^{(m)})} < (d_i+d_j-m-q)+ (m -d_j+q)=d_i$.
This shows that $\rho(f_i^{(m+1)})=d_i$ and $f_i^{(m+1)}$ is monic, 
as claimed in item~(\ref{ith1}).
By the induction hypothesis and Lemma~\ref{l:Hshift},
$x^pg_j^{(m)},x^p\psi_{j}^{(m)}$ satisfy the 
$d_j-q -pq= m-d_i$ key equation and $x^p(g_j^{(m)}S-\psi_{j}^{(m)})$ is
monic of order $q^2-q-1-(m-d_i)$.  Lemma~\ref{l:Hupdate} shows that
$f_i^{(m)}-\mu_i  x^pg_j^{(m)}, \varphi_i^{(m)} - \mu_i x^p \psi_j^{(m)}$
solves the $m+1-d_i$ key equation.  Since $d_i = \rho(f_i^{(m+1)})$, we
have established item~(\ref{ith2}) of the theorem.
Items~(\ref{ith3}) and~(\ref{ith4}) follow because
$g_i^{(m+1)}=g_i^{(m)}$ and 
$\psi_i^{(m+1)} = \psi_i^{(m)}$. 

Suppose now that $p<0$ and $\mu_i\neq 0$.
In this case, $f_i^{(m+1)}=x^{-p}f_i^{(m)}-\mu_i g_j^{(m)}$.
A simple computation shows
$\rho(x^{-p}f_i^{(m)})=m-d_j+q$
while $\rho{(g_j^{(m)})} < m-d_j+q$.
Thus,  $f_i^{(m+1)}$ is monic, and
\[\rho(f_i^{(m+1)})=\rho(x^{-p}f_i^{(m)})=m-d_j+q\equiv i\mod q.\]
From Lemma~\ref{l:Hshift},
$x^{-p}f_i^{(m)}, x^{-p}\varphi_i^{(m)}$ satisfy the  key equation of
 order $m-d_i +pq= d_j-q$.  By the induction hypothesis, $g_j^{(m)}$, $\psi_j^{(m)}$
satisfy the key equation of the same order.  Furthermore, $\mu_i$ is the coefficient of
$x^{-p-r_i-1}\zstar_j$ in $x^{-p}f_i^{(m)} S$.
Noting that $q(p+r_i ) +j(q+1) = d_j -q$ we may apply Lemma~\ref{l:Hupdate}
to obtain 
that $f_i^{(m+1)}$, $\varphi_i^{(m+1)}$ satisfy the
key equation of order $d_j-q+1 = m+1- \rho(f_i^{(m+1)} )$. This proves
item~(\ref{ith2}).

To prove items~(\ref{ith4}) and~(\ref{ith3}), we first establish
 that $\mu_j=\mu_i$.  We claim that each is the coefficient of
 $\zstar_0/x$ in $x^{-p-1}f_j^{(m)}f_i^{(m)}S$.
We know $\mu_i$ is the coefficient of $x^{-r_i-1}\zstar_j$ in $f_i^{(m)}S$,
which by Lemma~\ref{l:t_ab} is the coefficient of $\zstar_0/x$ in
$x^{r_i}y^jf_i^{(m)}S$.  Since $x^{-p-1}f_j^{(m)}$ and $x^{r_i}y^j$ are both monic
of order $m-d_i$, Lemma~\ref{l:gfS} says the coefficients of
$\zstar_0/x$ in $x^{r_i}y^jf_i^{(m)}S$ and
$x^{-p-1}f_j^{(m)}f_i^{(m)}S$ are equal.  A similar argument works for
$j$, which establishes the claim.

Since $\mu_j=\mu_i\ne 0$, the algorithm sets $g_i^{(m+1)}=\mu_i^{-1}f_j^{(m)}$
and $\psi_i^{(m)}=\mu_i^{-1}\varphi_j^{(m)}$.  
We can verify
item~(\ref{ith4}),
\begin{align*}
(m+1)-\rho(f_i^{(m+1)}) +q &= m+1-(m-d_{j}+q) +q \\
&= d_j +1 \\
&> \rho(g_i^{(m+1)})
\end{align*}
Item~(\ref{ith3}) also follows since 
$g_i^{(m+1)}, \psi_i^{(m+1)}$ satisfy the $m-d_ j$ 
key equation
and $m-d_j = \rho(f_i^{(m+1)})-q$.
Furthermore, $g_i^{(m+1)}S-\psi_i^{(m+1)} =\mu_j^{-1}(f_j^{(m)}S - \varphi_j^{(m)}) $ 
is monic.
\end{proof}

The next results establish the iteration number  $M$ at which the algorithm may be terminated.
This depends on the footprint of $e$,  $\Delta^e$, introduced earlier as well as the 
 orders of the Gr\"obner basis for $I^e$.
For $i=0$ up to $q-1$ define
$\sigma_i=\min\{\rho(f): f \in I^e \text{ and } \rho(f) \equiv i \mod q\}$.

\begin{lemma}
\label{lemma:rhof<sigma}
For all $m$ and for all $i$, 
$\rho{(f_i^{(m)})}\leq \sigma_i$.
\end{lemma}

\begin{proof}
Let $f_i^e \in I^e$ have pole order $\sigma_i$ and 
consider $f_i^e g_i^{(m)}S - f_i^e \psi_i^{(m)}$.
By Theorem~\ref{t:HBMS} (3), we  have
$\rho( f_i^e g^{(m)}S - f_i^e \psi^{(m)}) = \sigma_i+ q^2-1
-\rho(f_i^{(m)} )$.
This must be an element of $\Lam$ because $f_i^eS$, $g^{(m)}$ and
$\psi^{(m)}$  are all in  $\F_{q^2}[x,y]$.  
Since $\sigma_i-\rho(f_i^{(m)})$ is a multiple of $q$, and 
$q^2-q-1 \not \in \Lam$, we must have $\sigma_i-\rho(f_i^{(m)}) \geq 0$.
\end{proof}

\begin{proposition}
\label{p:HfinalM}
Let $\sigma_{\max}=\max\{\sigma_i:0\leq i\leq q-1\}$
and let $\del_{\max}=\max\{c \in \Delta^e\}$.
For  $m >\sigma_{\max}+\del_{\max}$, 
each of the polynomials $f_i^{(m)}$ belongs to $I^e$.
Let $M = \sigma_{\max}+ \max\{ \del_{\max}, q^2-q-1\}$.
Each of the pairs $f_i^{(M+1)}, \varphi_i^{(M+1)}$ satisfies the key equation.
\end{proposition}

\begin{proof}
By Theorem~\ref{t:HBMS}, $f_i^{(m)}, \varphi_i^{(m)}$ satisfy the 
$m-\rho(f_i^{(m)})$ key equation.  If $m >\sigma_{\max}+\del_{\max}$, 
then, $m-\rho(f_i^{(m)}) > \del_{\max}$,
so the result follows from Lemma~\ref{p:Hlocator characterization}.
For $ M = \sigma_{\max}+ \max\{ \del_{\max}, q^2-q-1\}$, we have
\[
\rho(f_i^{(M+1)}S-\varphi_i^{(M+1)} ) \leq q^2-q-1-(M +1 -
\rho(f_i^{(M+1)})) < 0.
\]
Since $f_i^{(M+1)}$ is a locator, $\varphi_i^{(M+1)}$ must equal $f_i^{(M+1)}S$.
\end{proof}

\subsection{Error evaluation without the error evaluator polynomials}

In this section we generalize the error evaluation formula in
Proposition~\ref{p:RSeval} that uses just the error locator polynomial $f$
and the update polynomial $g$ to determine error values.
The main result is Theorem~\ref{t:Hevalformula}, which is readily
derived from Proposition~\ref{p:sum dets}.   
Unfortunately, the proposition requires a result that takes some work  to 
establish: In the algorithm, when $i+j \equiv m \mod q$,  $\mu_i=\mu_j$.
This was shown for $p <0$ in the proof of Theorem~\ref{t:HBMS}, but in order to 
show it for $p\geq 0$ we need a rather technical result,
Proposition~\ref{p:OPdeterminants}. 
Since the result is easier to state using the language of
residues---instead of referring to the coefficient of $\zstar_0/x$---we
have deferred it to the section on general one-point codes.

\begin{proposition}
\label{p:sum dets}
Let $B_i^{(M)} = 
\begin{pmatrix}
f_i^{(m)}  &\varphi_i^{(m)}  \\
g_i^{(m)}  &\psi_i^{(m)}  \\
\end{pmatrix}$.  Then for all $m$, 
\begin{equation}
\label{e:det}
\sum_{i=0}^{q -1} \det B_i^{(m)} = -\sum_{i=0}^{q -1} y^i \zstar_i
=-1
\end{equation}
\end{proposition}

\begin{proof}
We proceed by induction.  The case $m=0$ is a simple calculation.
Assume that the statement of the theorem is true for $m$; we will
prove it for $m+1$.
It is sufficient to show that 
$\det B_i^{(m+1)}= \det B_i^{(m)}$ if $2i \equiv m \mod q$ and 
$\det B_i^{(m+1)}+ \det B_j^{(m+1)} = \det B_i^{(m)}+ \det B_j^{(m)}$ if
$i + j \equiv m \mod q$ and $i\not = j$.

If $2i \equiv m \mod q$, then $B_i^{(m+1)} = U_i^{(m)} B_i^{(m)}$,
where
\begin{align*}U_i^{(m)} &= 
 \begin{cases}
     \begin{pmatrix}1 & -\mu_i x^p\\
                      0 & 1
      \end{pmatrix} & \mbox{ if }\mu_i= 0 \mbox{ or }p\geq 0
\\
\begin{pmatrix}
x^{-p}&-\mu_i\\
1/\mu_i& 0
\end{pmatrix} & \mbox{ otherwise.}
\end{cases}
\end{align*}
Since $\det U_i^{(m)} = 1$ in either case, we have $\det B_i^{(m+1)}= \det B_i^{(m)}$.

Assume now that $i+j \equiv m \mod q$ and that $i\not=j$.  
Proposition~\ref{p:OPdeterminants} shows that $\mu_i=\mu_j$, so, from the algorithm 
\begin{align*}
B_i^{(m+1)} &= 
 \begin{cases}
     \begin{pmatrix}f_i^{(m)} - \mu_i x^p g_j^{(m)} & \varphi_i^{(m)} - \mu_i x^p \psi_j^{(m)}\\
                      g_i^{(m)} &  \psi_i^{(m)}
      \end{pmatrix} & \mbox{ if }\mu_i= 0 \mbox{ or }p\geq 0
\\
     \begin{pmatrix}x^{-p}f_i^{(m)} - \mu_i  g_j^{(m)} & x^{-p}\varphi_i^{(m)} - \mu_i  \psi_j^{(m)}\\
                      \mu_i^{-1}f_j^{(m)} &  \mu_i^{-1}\varphi_j^{(m)}
\end{pmatrix} & \mbox{ otherwise.}
\end{cases} \\
\intertext{The two cases lead respectively to }
\det B_i^{(m+1)} &= 
\begin{cases}
f_i^{(m)}\psi_i^{(m)} -g_i^{(m)}\varphi_i^{(m)} - 
             \mu_i x^p (g_j^{(m)}\psi_i^{(m)} -
             g_i^{(m)}\psi_j^{(m)}) & \text{ or} \\
f_j^{(m)}\psi_j^{(m)} -g_j^{(m)}\varphi_j^{(m)} - 
             \mu_i x^p (g_i^{(m)}\psi_j^{(m)} -
             g_j^{(m)}\psi_i^{(m)}) &  
\end{cases}
\end{align*}
To obtain $\det B_j^{(m+1)}$ one simply switches $i$ and $j$ in these
formulas.
When we take the sum of $\det B_i^{(m+1)}$ and  $\det B_j^{(m+1)}$, the
final terms cancel, so  $\det B_i^{(m+1)} +\det B_j^{(m+1)}
= \det B_i^{(m)} +\det B_j^{(m)}$.
\end{proof}

We now take $M$ as in Proposition~\ref{p:HfinalM}, so that the
algorithm of the previous section has produced 
solutions to the key equation.  Let
\begin{alignat*}{2}
f_i &= f_i^{(M+1)} &\quad \varphi_i &= \varphi_i^{(M+1)} \\
g_i &=g_i^{(M+1)} & \quad\psi_i &= \psi_i^{(M+1)} \\
\end{alignat*}
Then the $f_i$ are a basis for  $I^e$
 as a module over $\F_{q^2}[x,y]$ and $f_i, \varphi_i$ satisfy the key 
equation. 

\begin{theorem}
\label{t:Hevalformula}
If $P_k$ is an error position.
\begin{align}
\label{e:eval g}
e_k &=  \left(\sum_{i=0}^{q-1} f_i'(P_k)g_i(P_k) \right)^{-1}
\end{align}
\end{theorem}

\begin{proof}
From the preceding lemma, 
\[\sum_{i=0}^{q-1} \left( f_i \psi_i - g_i\varphi_i \right) = -1
\]
Evaluating at an error position $P_k$  we have
$-\sum_{i=0}^{q-1}g_i\varphi_i(P_k) =- 1$.  
Apply Lemma~\ref{l:Hevaluator}, to get
$\sum_{i=0}^{q-1} g_i(P_k) e_k f_i'(P_k) = 1$.
Solving for $e_k$  gives the formula.
\end{proof}

\subsection{An example}
Consider the Hermitian curve associated to the field extension
$\F_9 = \F_3[\alpha]$ where  $\alpha^2=\alpha+1$.
Let $x$ and $y$ be the classes of $X$ and $Y$ in the quotient
${\mathbb F}_9[X,Y]/(X^4-Y^3-Y)$.  The basis monomials are $x^ay^b$
for $a\geq0 $ and $0 \leq b \leq 2$.
The order of $x$ is $3$ and the order of $y$ is $4$.
So, \[\Lambda=\{ 0, 3, 4, 6 ,7,8,9,10,\dots\}.\]

The Hermitian curve in this case has 27 points, which we take in the
following order:
$   ( 1, \alpha )$,
$    ( 1, \alpha^3 )$,
$    ( 1, 2 )$,
$    ( \alpha, 1 )$,
$    ( \alpha, \alpha^5 )$,
$    ( \alpha, \alpha^7 )$,
$    ( \alpha^2, \alpha )$,
$    ( \alpha^2, \alpha^3 )$,
$    ( \alpha^2, 2 )$,
$    ( \alpha^3, 1 )$,
$    ( \alpha^3, \alpha^5 )$,
$    ( \alpha^3, \alpha^7 )$,
$    ( 2, \alpha )$,
$    ( 2, \alpha^3 )$,
$    ( 2, 2 )$,
$    ( \alpha^5, 1 )$,
$    ( \alpha^5, \alpha^5 )$,
$   ( \alpha^5, \alpha^7 )$,
$    ( \alpha^6, \alpha )$,
$    ( \alpha^6, \alpha^3 )$,
$    ( \alpha^6, 2 )$,
$    ( \alpha^7, 1 )$,
$    ( \alpha^7, \alpha^5 )$,
$    ( \alpha^7, \alpha^7 )$,
$    ( 0, \alpha^2 )$,
$    ( 0, \alpha^6 )$,
$    ( 0, 0 )$.

Let us consider correction of two errors.
There are two choices for $\Delta^e$ when the weight of $e$ is two, 
$\{0,4\}$ when the points are on a vertical line, and $\{0,3\}$ when
they are not.  Following~\cite{BrasOS:Correct}, we will call the
latter case ``generic'' and former ``non-generic.''  In either case
$\sigma_{\max}=8$.   
From Proposition~\ref{p:HfinalM}, the computation of all error
locators and evaluators is complete after  iteration number
$M=\sigma_{\max}+\max\{\delta_{\max}, q^2-q-1\}$.  
Thus, for an algorithm to correct either of the two errors we
terminate the algorithm  with iteration $M=8+5=13$, and take the data
for superscript $14$.
We will explain in detail the first steps in the generic case. 
All the computations  are summarized in Table~\ref{tab:generic}.
The computations in the non-generic case are summarized in
Table~\ref{tab:non-generic}. 

For the generic error vector we take error values
$\alpha^2$ at the point $(\alpha,1)$ and $\alpha^7$ at the point
$(\alpha^6, \alpha^3)$, so the error vector is 
\[e = (  0   0   0 \alpha^2   0   0   0   0   0   0   0   0   0   0   0
0   0   0   0   0 \alpha^7   0   0   0   0   0   0).\]
The associated syndromes are 
$$\begin{array}{c|ccccccccc}
&s_{ 0 ,b}
&s_{ 1 ,b}
&s_{ 2 ,b}
&s_{ 3 ,b}
&s_{ 4 ,b}
&s_{ 5 ,b}
&s_{ 6 ,b}
&s_{ 7 ,b}
&s_{ 8 ,b}
\\ \hline
s_{a, 0 }
&
\alpha^5
&
\alpha^2
&
\alpha^5
&
0
&
1
&
2
&
\alpha^6
&
\alpha^5
&
\alpha^5
\\ 
s_{a, 1 }
&
2
&
1
&
\alpha^2
&
\alpha
&
\alpha
&
\alpha^6
&
\alpha
&
0
&
2
\\ 
s_{a, 2 }
&
\alpha^5
&
\alpha^2
&
\alpha^5
&
0
&
1
&
2
&
\alpha^6
&
\alpha^5
&
\alpha^5
\\ 
\end{array}$$

To initialize $f$, $g$, $\varphi$, $\psi$ we take \newline
\hspace*{1in}\begin{tabular}{llll}
$f_0=1$ &  $g_0=0$ & $\varphi_0=0 $ & $\psi_0=2y^2+2$ \\
$f_1=y$ & $g_1=0$ &  $ \varphi_1= 0$ &  $\psi_1=2y$ \\
$g_2=0$ & $f_2=y^2$ & $\varphi_2=0$ &    $\psi_2=2$
\end{tabular}

We start with $m=0$.
The pairs $i,j$ with $i+j\equiv m \mbox{ mod }3$ are
$0,0$ and $1,2$.
The data computed in the algorithm is, \newline
\hspace*{1in}\begin{tabular}{lll}
$r_0=0 $& $\tilde{f}_0=1 $&$ \mu_0=s_{0,0}=\alpha^5$,\\
$r_1=-4 $& $\tilde{f}_1=x^4+2y $&$  \mu_1=s_{0,0}+2s_{-4,1}=\alpha^5$,\\
$r_2=-4$ &$ \tilde{f}_2=x^4+2y $& $ \mu_2=s_{0,0}+2s_{-4,1}=\alpha^5$.\\
\end{tabular}\\
For the pair $0,0$, $p=-1$, and for the pair $1,2$, $p=3$,  so
\[
U_0^{(0)}=\left(\begin{array}{cc}x&\alpha\\\alpha^3&0\end{array}\right)
\qquad \text{ and } \qquad U_1^{(0)}=U_2^{(0)}=
\left(\begin{array}{cc}1&\alpha x^3\\0&1\end{array}\right).
\]
As a result,\newline
\hspace*{1in}\begin{tabular}{llll}
$f_0^{(1)}=x$ &  $\varphi_0^{(1)}=\alpha^5y^2+\alpha^5$ &  $g_0^{(1)}=\alpha^3$ &  $\psi_0^{(1)}=0$,
\\
 $f_1^{(1)}=y$ &  $\varphi_1^{(1)}=\alpha^5x^3$ &   $g_1^{(1)}=0$ &  $\psi_1^{(1)}=2y$ \\
$f_2^{(1)}=y^2$ &  $\varphi_2^{(1)}=\alpha^5x^3y$ &  $g_2^{(1)}=0$ &  $\psi_2^{(1)}=2$
\end{tabular}

For $m=1$, 
the pairs $i,j$ with $i+j\equiv m \mbox{ mod }3$ are
$0,1$ and $2,2$, and,  
\hspace*{1in}\begin{tabular}{lll}
$r_0=-2 $ & $\tilde{f}_0=xy $ & $\mu_0=s_{-1,1}=0$,\\
$r_1=-1 $ & $\tilde{f}_1=y $ & $ \mu_1=s_{-1,1}=0$,\\
$r_2=-5 $ & $\tilde{f}_2=x^4y+2y^2 $ & $ \mu_2=s_{-1,1}+2s_{-5,2}=0$.\\
\end{tabular}\\
This means that
\[U_0^{(1)}=U_1^{(1)}=U_2^{(1)}=\left(\begin{array}{cc}1&0\\0&1\end{array}\right)\]
and $f_i, \varphi_i, g_i, \psi_i$ remain unchanged.

For $m=2$, 
the pairs $i,j$ with $i+j\equiv m \mbox{ mod }3$ are
$0,2$ and $1,1$, and,\\
\hspace*{1in}\begin{tabular}{lll}
$r_0=-3$ & $\tilde{f}_0=xy^2$ & $\mu_0=s_{-2,2}=0$\\
$r_1=-2$ & $\tilde{f}_1=y^2$ & $ \mu_1=s_{-2,2}=0$\\
$r_2=-2$ & $\tilde{f}_2=y^2$ & $ \mu_2=s_{-2,2}+2s_{-2,2}=0$\\
\end{tabular}\\
Again, this means 
$U_0^{(2)}=U_1^{(2)}=U_2^{(2)}=\left(\begin{array}{cc}1&0\\0&1\end{array}\right)$,
and $f_i, \varphi_i, g_i, \psi_i$ remain unchanged.

For $m=3$, 
the pairs $i,j$ with $i+j\equiv m \mbox{ mod }3$ are
$0,0$ and $1,2$. Now, \\
\hspace*{1in}\begin{tabular}{lll}
$r_0=0$ & $\tilde{f}_0=x$ & $\mu_0=s_{1,0}=\alpha^2$\\
$r_1=-3$ & $\tilde{f}_1=x^4+2y$ & $ \mu_1=s_{1,0}+2s_{-3,1}=\alpha^2$\\
$r_2=-3$ & $\tilde{f}_2=x^4+2y$ & $ \mu_2=s_{1,0}+2s_{-3,1}=\alpha^2$\\
\end{tabular}\\
For the pair $0,0$ we have $p=0$ and 
for the pair $1,2$ we have $p=2$ so,
\[U_0^{(3)}=\left(\begin{array}{cc}1&\alpha^6\\0&1\end{array}\right)
\qquad \text{ and } \qquad U_1^{(3)}=U_2^{(3)}=
\left(\begin{array}{cc}1&\alpha^6 x^2\\0&1\end{array}\right)
\]
Consequently, \newline
\hspace*{.3in}\begin{tabular}{llll}
$f_0^{(4)}=x+\alpha$ & $\varphi_0^{(4)}=\alpha^5y^2+\alpha^5$ &
$g_0^{(4)}=\alpha^3$   &$\psi_0^{(4)}=0$  \\
 $f_1^{(4)}=y$ &    $\varphi_1^{(4)}=\alpha^5x^3+\alpha^2x^2$ & 
$g_1^{(4)}=0$ & $\psi_1^{(4)}=2y$\\
  $f_2^{(4)}=y^2$   &  $\varphi_2^{(4)}=\alpha^5x^3y+\alpha^2x^2y$ &
$g_2^{(4)}=0$ &   $\psi_2^{(4)}=2$
\end{tabular}

The subsequent steps are summarized in  Table~\ref{tab:generic}.
Let $f_i= f_i^{(14)}$ and similarly for the other data.  
The locators and associated derivatives are (using
$\frac{dy}{dx}=x^3$), \\
\hspace*{0.5in}\begin{tabular}{ll}
$f_0=x^2 + x + \alpha^7$ &  $(f_0)' = 2x+1$\\
$f_1=y + \alpha^5x + \alpha$ &$\left(f_1\right)'=x^3+\alpha^5$\\
$f_2=y^2 + \alpha^7x^2 + \alpha^7x + \alpha^3$ &
$(f_2)'=2 x^3 y + \alpha^3 x + \alpha^7$
\end{tabular}

The points where $f_0$, $f_1$, and $f_2$ vanish 
are exactly $P_4=(\alpha,1)$ and $P_{21}=(\alpha^6,2)$,
coinciding with the error positions.
The polynomials $\varphi$ are:
\\
\hspace*{0.25in}\begin{tabular}{l}
$\varphi_0=\alpha^5xy^2 + y^2 + 2xy + \alpha x + 2$,\\
$\varphi_1=\alpha^5x^3 + \alpha^2y^2 + \alpha^2x^2 + \alpha y + \alpha^5x + a^6$,\\
$\varphi_2=\alpha^5 x^3 y + 2 x y^2 + \alpha^2 x^2 y + 2 x^3 + \alpha^7 y^2
+ \alpha^2 x y + x^2 + \alpha^7 x + 1$
\end{tabular}\\
The error values at these positions can be computed using 
the formula in Lemma~\ref{l:Hevaluator}. 
For example,
\begin{align*}
e_4&=\frac{\varphi_1(P_4)}{\left(f_1\right)'(P_{4})}=
\frac{\alpha^5}{\alpha^3}=\alpha^2
 & \quad \quad e_{21}&=\frac{\varphi_1(P_{21})}{\left(f_1\right)'(P_{21})}=
\frac{\alpha^6}{\alpha^7}=\alpha^7.
\end{align*}
The same error values could have been obtained using
$f_0$ and $\varphi_0$
instead of $f_1$ and $\varphi_1$.
However, we could not have used 
$f_2$ 
and $\varphi_2$,
because the zero of 
$f_2$ 
at $P_{21}$ is not simple.

By Theorem~\ref{t:Hevalformula}, the error values can also be obtained
using $g_0,g_1,g_2$ instead of $\varphi_0,\varphi_1,\varphi_2$.
Since $g_0=\alpha^6x+\alpha^7$, and $g_1=g_2=0$ there is only one term
to compute.
\begin{alignat*}{2}
e_4 &= \left(f_0'(P_4)g_0(P_4)\right)^{-1} & \qquad \qquad 
   e_{21} &= \left(\left(f_0\right)'(P_{21})g_0(P_{21})\right)^{-1} \\
&=\left(\alpha^3\cdot\alpha^3 \right)^{-1} & \qquad \qquad 
  &=\left( \alpha^7\cdot\alpha^2 \right)^{-1}\\
&=\alpha^2 & \qquad \qquad &=\alpha^7.
\end{alignat*}

An example of a non-generic  error vector is 
\[(
0 0 0 0 0 0 \alpha^2 0 \alpha^7
0 0 0 0 0 0 0 0 0
0 0 0 0 0 0 0 0 0
).\]
The error positions correspond to the points
$P_7=(\alpha^2,\alpha)$ and $P_9(\alpha^2,2)$
which lie on the line $x=\alpha^2$.
The associated syndromes are
$$\begin{array}{c|ccccccccc}
&s_{ 0 ,c}
&s_{ 1 ,c}
&s_{ 2 ,c}
&s_{ 3 ,c}
&s_{ 4 ,c}
&s_{ 5 ,c}
&s_{ 6 ,c}
&s_{ 7 ,c}
&s_{ 8 ,c}
\\ \hline
s_{a, 0 }
&
\alpha^5
&
\alpha^7
&
\alpha
&
\alpha^3
&
\alpha^5
&
\alpha^7
&
\alpha
&
\alpha^3
&
\alpha^5
\\ 
s_{a, 1 }
&
\alpha^7
&
\alpha
&
\alpha^3
&
\alpha^5
&
\alpha^7
&
\alpha
&
\alpha^3
&
\alpha^5
&
\alpha^7
\\ 
s_{a, 2 }
&
\alpha^2
&
2
&
\alpha^6
&
1
&
\alpha^2
&
2
&
\alpha^6
&
1
&
\alpha^2
\\ 
\end{array}$$
The steps of the algorithm are summarized in Table~\ref{tab:non-generic}.
Notice that after step $m=4$, $f_0=x-\alpha^2=x+\alpha^6$ is already a
locator.

\begin{table}[h]
\label{tab:generic}
\resizebox{\textwidth}{!}{\rotatebox{90}{{\tiny$\begin{array}{|r|llrllr|l|l|l|l|l|}
\hline {\bf m} & {\bf i} & {\bf j} & {\bf r_i} &{\bf \tilde{f}_i} & {\bf \mu_i} 
& {\bf p} & {\bf U_i^{(m)}} & {\bf f_i^{(m+1)}}  & {\bf \varphi_i^{(m+1)}} & 
{\bf g_i^{(m+1)}} & {\bf \psi_i^{(m+1)}}\\ \hline
-1 & 0 &&&&&&& 1
& 0
& 0
& 2 y^2 + 2
\\
& 1 &&&&&&& y
& 0
& 0
& 2 y
\\
& 2 &&&&&&& y^2
& 0
& 0
& 2
\\
\hline  0
& 0 & 0 & 0 & 1
& \alpha^5 & -1
& [
    [
        x,
        \alpha
    ],
    [
        \alpha^3,
        0
    ]
]
& x
& \alpha^5 y^2 + \alpha^5
& \alpha^3
& 0
\\ & 1 & 2 & -4 & x^4 + 2 y
& \alpha^5 & 3
& [
    [
        1,
        \alpha x^3
    ],
    [
        0,
        1
    ]
]
& y
&\begin{array}[t]{l} \alpha^5 x^3
\\       \end{array}&&
\\ & 2 & 1 & -4 & x^4 + 2 y
& \alpha^5 & 3
& [
    [
        1,
        \alpha x^3
    ],
    [
        0,
        1
    ]
]
& y^2
&\begin{array}[t]{l} \alpha^5 x^3 y
\\       \end{array}&&
\\ \hline  1
& 0 & 1 & -2 & x y
& 0 & 1
&&&&&
\\ & 1 & 0 & -1 & y
& 0 & 1
&&&&&
\\ & 2 & 2 & -5 & x^4 y + 2 y^2
& 0 & 4
&&&&&
\\ \hline  2
& 0 & 2 & -3 & x y^2
& 0 & 2
&&&&&
\\ & 1 & 1 & -2 & y^2
& 0 & 1
&&&&&
\\ & 2 & 0 & -2 & y^2
& 0 & 2
&&&&&
\\ \hline  3
& 0 & 0 & 0 & x
& \alpha^2 & 0
& [
    [
        1,
        \alpha^6
    ],
    [
        0,
        1
    ]
]
& x + \alpha
&\begin{array}[t]{l} \alpha^5 y^2 + \alpha^5
\\       \end{array}&&
\\ & 1 & 2 & -3 & x^4 + 2 y
& \alpha^2 & 2
& [
    [
        1,
        \alpha^6 x^2
    ],
    [
        0,
        1
    ]
]
& y
&\begin{array}[t]{l} \alpha^5 x^3 + \alpha^2 x^2
\\       \end{array}&&
\\ & 2 & 1 & -3 & x^4 + 2 y
& \alpha^2 & 2
& [
    [
        1,
        \alpha^6 x^2
    ],
    [
        0,
        1
    ]
]
& y^2
&\begin{array}[t]{l} \alpha^5 x^3 y + \alpha^2 x^2 y
\\       \end{array}&&
\\ \hline  4
& 0 & 1 & -1 & x y + \alpha y
& 2 & 0
& [
    [
        1,
        1
    ],
    [
        0,
        1
    ]
]
& x + \alpha
&\begin{array}[t]{l} \alpha^5 y^2 + 2 y + \alpha^5
\\       \end{array}&&
\\ & 1 & 0 & 0 & y
& 2 & 0
& [
    [
        1,
        1
    ],
    [
        0,
        1
    ]
]
& y + \alpha^3
&\begin{array}[t]{l} \alpha^5 x^3 + \alpha^2 x^2
\\       \end{array}&&
\\ & 2 & 2 & -4 & x^4 y + 2 y^2
& 2 & 3
& [
    [
        1,
        x^3
    ],
    [
        0,
        1
    ]
]
& y^2
&\begin{array}[t]{l} \alpha^5 x^3 y + \alpha^2 x^2 y + 2 x^3
\\       \end{array}&&
\\ \hline  5
& 0 & 2 & -2 & x y^2 + \alpha y^2
& 0 & 1
&&&&&
\\ & 1 & 1 & -1 & y^2 + \alpha^3 y
& 0 & 0
&&&&&
\\ & 2 & 0 & -1 & y^2
& 0 & 1
&&&&&
\\ \hline  6
& 0 & 0 & 1 & x + \alpha
& \alpha^2 & -1
& [
    [
        x,
        \alpha^6
    ],
    [
        \alpha^6,
        0
    ]
]
& x^2 + \alpha x + \alpha
& \alpha^5 x y^2 + 2 x y + \alpha^5 x
& \alpha^6 x + \alpha^7
& \alpha^3 y^2 + \alpha^2 y + \alpha^3
\\ & 1 & 2 & -2 & x^4 + \alpha^3 y^2 + 2 y
& \alpha^5 & 1
& [
    [
        1,
        \alpha x
    ],
    [
        0,
        1
    ]
]
& y + \alpha^3
&\begin{array}[t]{l} \alpha^5 x^3 + \alpha^2 x^2 + \alpha^5 x
\\       \end{array}&&
\\ & 2 & 1 & -2 & x^4 + 2 y
& \alpha^5 & 1
& [
    [
        1,
        \alpha x
    ],
    [
        0,
        1
    ]
]
& y^2
&\begin{array}[t]{l} \alpha^5 x^3 y + \alpha^2 x^2 y + 2 x^3 + 
    \alpha^5 x y
\\       \end{array}&&
\\ \hline  7
& 0 & 1 & -1 & x^2 y + \alpha x y + \alpha y
& \alpha^3 & 0
& [
    [
        1,
        \alpha^7
    ],
    [
        0,
        1
    ]
]
& x^2 + \alpha x + \alpha
&\begin{array}[t]{l} \alpha^5 x y^2 + 2 x y + \alpha^3 y + \alpha^5 x
\\       \end{array}&&
\\ & 1 & 0 & 1 & y + \alpha^3
& \alpha^3 & 0
& [
    [
        1,
        \alpha^7
    ],
    [
        0,
        1
    ]
]
& y + \alpha^5 x + \alpha
&\begin{array}[t]{l} \alpha^5 x^3 + \alpha^2 y^2 + \alpha^2 x^2 + 
    \alpha y + \alpha^5 x
+ \alpha^2
\end{array}&&
\\ & 2 & 2 & -3 & x^4 y + 2 y^2
& 1 & 2
& [
    [
        1,
        2 x^2
    ],
    [
        0,
        1
    ]
]
& y^2
&\begin{array}[t]{l} \alpha^5 x^3 y + \alpha^2 x^2 y + 2 x^3 + 
    \alpha^5 x y + x^2
\\       \end{array}&&
\\ \hline  8
& 0 & 2 & -2 & x^2 y^2 + \alpha x y^2 + \alpha y^2
& \alpha^5 & 1
& [
    [
        1,
        \alpha x
    ],
    [
        0,
        1
    ]
]
& x^2 + \alpha x + \alpha
&\begin{array}[t]{l} \alpha^5 x y^2 + 2 x y + \alpha^3 y + \alpha x
\\       \end{array}&&
\\ & 1 & 1 & 0 & y^2 + \alpha^5 x y + \alpha y
& 0 & -1
&&&&&
\\ & 2 & 0 & 0 & y^2
& \alpha^5 & 1
& [
    [
        1,
        \alpha x
    ],
    [
        0,
        1
    ]
]
& y^2 + \alpha^7 x^2 + x
&\begin{array}[t]{l} \alpha^5 x^3 y + 2 x y^2 + \alpha^2 x^2 y + 2 x^3 + 
    \alpha^2 x y
\\ \phantom{mmmmmmmm}\hfill{+ x^2 + 2 x
} \end{array}&&
\\ \hline  9
& 0 & 0 & 1 & x^2 + \alpha x + \alpha
& \alpha & 0
& [
    [
        1,
        \alpha^5
    ],
    [
        0,
        1
    ]
]
& x^2 + x + \alpha^7
&\begin{array}[t]{l} \alpha^5 x y^2 + y^2 + 2 x y + \alpha x + 1
\\       \end{array}&&
\\ & 1 & 2 & -1 & x^4 + \alpha^5 x y^2 + \alpha y^2 + 2 y
& \alpha^2 & 0
& [
    [
        1,
        \alpha^6
    ],
    [
        0,
        1
    ]
]
& y + \alpha^5 x + \alpha
&\begin{array}[t]{l} \alpha^5 x^3 + \alpha^2 y^2 + \alpha^2 x^2 + 
    \alpha y + \alpha^5 x
\\ \phantom{mmmmmmmm}\hfill{+ \alpha^6
} \end{array}&&
\\ & 2 & 1 & -1 & x^4 + \alpha^7 x^2 y + x y + 2 y
& \alpha^2 & 0
& [
    [
        1,
        \alpha^6
    ],
    [
        0,
        1
    ]
]
& y^2 + \alpha^7 x^2 + x
&\begin{array}[t]{l} \alpha^5 x^3 y + 2 x y^2 + \alpha^2 x^2 y + 2 x^3 + 
    \alpha^2 x y
\\ \phantom{mmmmmmmm}\hfill{+ x^2 + \alpha^2 y + 2 x
} \end{array}&&
\\ \hline  10
& 0 & 1 & 0 & x^2 y + x y + \alpha^7 y
& 0 & -1
&&&&&
\\ & 1 & 0 & 2 & y + \alpha^5 x + \alpha
& 0 & -1
&&&&&
\\ & 2 & 2 & -2 & x^4 y + \alpha^7 x^2 y^2 + x y^2 + 2 y^2
& \alpha & 1
& [
    [
        1,
        \alpha^5 x
    ],
    [
        0,
        1
    ]
]
& y^2 + \alpha^7 x^2 + x
&\begin{array}[t]{l} \alpha^5 x^3 y + 2 x y^2 + \alpha^2 x^2 y + 2 x^3 + 
    \alpha^2 x y
\\ \phantom{mmmmmmmm}\hfill{+ x^2 + \alpha^2 y + \alpha^7 x
} \end{array}&&
\\ \hline  11
& 0 & 2 & -1 & x^2 y^2 + x y^2 + \alpha^7 y^2
& 1 & 0
& [
    [
        1,
        2
    ],
    [
        0,
        1
    ]
]
& x^2 + x + \alpha^7
&\begin{array}[t]{l} \alpha^5 x y^2 + y^2 + 2 x y + \alpha x + 2
\\       \end{array}&&
\\ & 1 & 1 & 1 & y^2 + \alpha^5 x y + \alpha y
& 0 & -2
&&&&&
\\ & 2 & 0 & 1 & y^2 + \alpha^7 x^2 + x
& 1 & 0
& [
    [
        1,
        2
    ],
    [
        0,
        1
    ]
]
& y^2 + \alpha^7 x^2 + \alpha^7 x + \alpha^3
&\begin{array}[t]{l} \alpha^5 x^3 y + 2 x y^2 + \alpha^2 x^2 y + 2 x^3 + 
    \alpha^7 y^2
\\ \phantom{mmmmmmmm}\hfill{+ \alpha^2 x y + x^2 + \alpha^7 x + 
    \alpha^7
} \end{array}&&
\\ \hline  12
& 0 & 0 & 2 & x^2 + x + \alpha^7
& 0 & -1
&&&&&
\\ & 1 & 2 & 0 & x^4 + \alpha^5 x y^2 + \alpha y^2 + 2 y
& 0 & -1
&&&&&
\\ & 2 & 1 & 0 & x^4 + \alpha^7 x^2 y + \alpha^7 x y + \alpha^5 y
& 0 & -1
&&&&&
\\ \hline  13
& 0 & 1 & 1 & x^2 y + x y + \alpha^7 y
& 0 & -2
&&&&&
\\ & 1 & 0 & 3 & y + \alpha^5 x + \alpha
& 0 & -2
&&&&&
\\ & 2 & 2 & -1 & x^4 y + \alpha^7 x^2 y^2 + \alpha^7 x y^2 + \alpha^5 y^2
& \alpha^6 & 0
& [
    [
        1,
        \alpha^2
    ],
    [
        0,
        1
    ]
]
& y^2 + \alpha^7 x^2 + \alpha^7 x + \alpha^3
&\begin{array}[t]{l} \alpha^5 x^3 y + 2 x y^2 + \alpha^2 x^2 y + 2 x^3 + \alpha^7 y^2
\\ \phantom{mmmmmmmm}\hfill{+ \alpha^2 x y + x^2 + \alpha^7 x + 1
} \end{array}&&
\\ \hline\end{array}$}}}

\caption{
Steps for correcting two errors in general position.
}
\end{table}
\begin{table}[h]
\label{tab:non-generic}
\resizebox{!}{\textheight}{\rotatebox{90}{{\tiny$\begin{array}{|r|llrllr|l|l|l|l|l|}
\hline {\bf m} & {\bf i} & {\bf j} & {\bf r_i} &{\bf \tilde{f}_i} & {\bf \mu_i} 
& {\bf p} & {\bf U_i^{(m)}} & {\bf f_i^{(m+1)}}  & {\bf \varphi_i^{(m+1)}} & 
{\bf g_i^{(m+1)}} & {\bf \psi_i^{(m+1)}}\\ \hline
-1 & 0 &&&&&&& 1
& 0
& 0
& 2 y^2 + 2
\\
& 1 &&&&&&& y
& 0
& 0
& 2 y
\\
& 2 &&&&&&& y^2
& 0
& 0
& 2
\\
\hline  0
& 0 & 0 & 0 & 1
& \alpha^5 & -1
& [
    [
        x,
        \alpha
    ],
    [
        \alpha^3,
        0
    ]
]
& x
& \alpha^5 y^2 + \alpha^5
& \alpha^3
& 0
\\ & 1 & 2 & -4 & x^4 + 2 y
& \alpha^5 & 3
& [
    [
        1,
        \alpha x^3
    ],
    [
        0,
        1
    ]
]
& y
&\begin{array}[t]{l} \alpha^5 x^3
\\       \end{array}&&
\\ & 2 & 1 & -4 & x^4 + 2 y
& \alpha^5 & 3
& [
    [
        1,
        \alpha x^3
    ],
    [
        0,
        1
    ]
]
& y^2
&\begin{array}[t]{l} \alpha^5 x^3 y
\\       \end{array}&&
\\ \hline  1
& 0 & 1 & -2 & x y
& 0 & 1
&&&&&
\\ & 1 & 0 & -1 & y
& 0 & 1
&&&&&
\\ & 2 & 2 & -5 & x^4 y + 2 y^2
& 0 & 4
&&&&&
\\ \hline  2
& 0 & 2 & -3 & x y^2
& 0 & 2
&&&&&
\\ & 1 & 1 & -2 & y^2
& 0 & 1
&&&&&
\\ & 2 & 0 & -2 & y^2
& 0 & 2
&&&&&
\\ \hline  3
& 0 & 0 & 0 & x
& \alpha^7 & 0
& [
    [
        1,
        \alpha^3
    ],
    [
        0,
        1
    ]
]
& x + \alpha^6
&\begin{array}[t]{l} \alpha^5 y^2 + \alpha^5
\\       \end{array}&&
\\ & 1 & 2 & -3 & x^4 + 2 y
& \alpha^7 & 2
& [
    [
        1,
        \alpha^3 x^2
    ],
    [
        0,
        1
    ]
]
& y
&\begin{array}[t]{l} \alpha^5 x^3 + \alpha^7 x^2
\\       \end{array}&&
\\ & 2 & 1 & -3 & x^4 + 2 y
& \alpha^7 & 2
& [
    [
        1,
        \alpha^3 x^2
    ],
    [
        0,
        1
    ]
]
& y^2
&\begin{array}[t]{l} \alpha^5 x^3 y + \alpha^7 x^2 y
\\       \end{array}&&
\\ \hline  4
& 0 & 1 & -1 & x y + \alpha^6 y
& \alpha^7 & 0
& [
    [
        1,
        \alpha^3
    ],
    [
        0,
        1
    ]
]
& x + \alpha^6
&\begin{array}[t]{l} \alpha^5 y^2 + \alpha^7 y + \alpha^5
\\       \end{array}&&
\\ & 1 & 0 & 0 & y
& \alpha^7 & 0
& [
    [
        1,
        \alpha^3
    ],
    [
        0,
        1
    ]
]
& y + \alpha^6
&\begin{array}[t]{l} \alpha^5 x^3 + \alpha^7 x^2
\\       \end{array}&&
\\ & 2 & 2 & -4 & x^4 y + 2 y^2
& \alpha^7 & 3
& [
    [
        1,
        \alpha^3 x^3
    ],
    [
        0,
        1
    ]
]
& y^2
&\begin{array}[t]{l} \alpha^5 x^3 y + \alpha^7 x^2 y + \alpha^7 x^3
\\       \end{array}&&
\\ \hline  5
& 0 & 2 & -2 & x y^2 + \alpha^6 y^2
& 0 & 1
&&&&&
\\ & 1 & 1 & -1 & y^2 + \alpha^6 y
& 0 & 0
&&&&&
\\ & 2 & 0 & -1 & y^2
& 0 & 1
&&&&&
\\ \hline  6
& 0 & 0 & 1 & x + \alpha^6
& 0 & -1
&&&&&
\\ & 1 & 2 & -2 & x^4 + \alpha^6 y^2 + 2 y
& \alpha & 1
& [
    [
        1,
        \alpha^5 x
    ],
    [
        0,
        1
    ]
]
& y + \alpha^6
&\begin{array}[t]{l} \alpha^5 x^3 + \alpha^7 x^2 + \alpha x
\\       \end{array}&&
\\ & 2 & 1 & -2 & x^4 + 2 y
& \alpha & 1
& [
    [
        1,
        \alpha^5 x
    ],
    [
        0,
        1
    ]
]
& y^2
&\begin{array}[t]{l} \alpha^5 x^3 y + \alpha^7 x^2 y + \alpha^7 x^3 + 
    \alpha x y
\\       \end{array}&&
\\ \hline  7
& 0 & 1 & 0 & x y + \alpha^6 y
& 0 & -1
&&&&&
\\ & 1 & 0 & 1 & y + \alpha^6
& 0 & -1
&&&&&
\\ & 2 & 2 & -3 & x^4 y + 2 y^2
& \alpha & 2
& [
    [
        1,
        \alpha^5 x^2
    ],
    [
        0,
        1
    ]
]
& y^2
&\begin{array}[t]{l} \alpha^5 x^3 y + \alpha^7 x^2 y + \alpha^7 x^3 + 
    \alpha x y + \alpha x^2
\\       \end{array}&&
\\ \hline  8
& 0 & 2 & -1 & x y^2 + \alpha^6 y^2
& \alpha^2 & 0
& [
    [
        1,
        \alpha^6
    ],
    [
        0,
        1
    ]
]
& x + \alpha^6
&\begin{array}[t]{l} \alpha^5 y^2 + \alpha^7 y + 1
\\       \end{array}&&
\\ & 1 & 1 & 0 & y^2 + \alpha^6 y
& 1 & -1
& [
    [
        x,
        2
    ],
    [
        1,
        0
    ]
]
& x y + \alpha^6 x
& \alpha^5 x^4 + \alpha^7 x^3 + \alpha x^2 + y
& y + \alpha^6
& \alpha^5 x^3 + \alpha^7 x^2 + \alpha x
\\ & 2 & 0 & 0 & y^2
& \alpha^2 & 0
& [
    [
        1,
        \alpha^6
    ],
    [
        0,
        1
    ]
]
& y^2 + \alpha
&\begin{array}[t]{l} \alpha^5 x^3 y + \alpha^7 x^2 y + \alpha^7 x^3 + 
    \alpha x y + \alpha x^2
\\       \end{array}&&
\\ \hline  9
& 0 & 0 & 2 & x + \alpha^6
& 0 & -2
&&&&&
\\ & 1 & 2 & -2 & x^5 + \alpha^6 x y^2 + 2 x y
& \alpha^3 & 1
& [
    [
        1,
        \alpha^7 x
    ],
    [
        0,
        1
    ]
]
& x y + \alpha^6 x
&\begin{array}[t]{l} \alpha^5 x^4 + \alpha^7 x^3 + \alpha x^2 + y + 
    \alpha^3 x
\\       \end{array}&&
\\ & 2 & 1 & -1 & x^4 + \alpha^7 y
& \alpha^3 & 1
& [
    [
        1,
        \alpha^7 x
    ],
    [
        0,
        1
    ]
]
& y^2 + \alpha^7 x y + \alpha^5 x + \alpha
&\begin{array}[t]{l} \alpha^5 x^3 y + 2 x^4 + \alpha^7 x^2 y + x^3 + 
    \alpha x y
\\ \phantom{mmmmmmmm}\hfill{+ \alpha^2 x^2
} \end{array}&&
\\ \hline  10
& 0 & 1 & 1 & x y + \alpha^6 y
& 0 & -1
&&&&&
\\ & 1 & 0 & 1 & x y + \alpha^6 x
& 0 & -1
&&&&&
\\ & 2 & 2 & -2 & x^4 y + \alpha^7 x^5 + \alpha^5 x y^2 + \alpha^7 y^2 
    + \alpha^3 x y
& 2 & 1
& [
    [
        1,
        x
    ],
    [
        0,
        1
    ]
]
& y^2 + \alpha^7 x y + \alpha^5 x + \alpha
&\begin{array}[t]{l} \alpha^5 x^3 y + 2 x^4 + \alpha^7 x^2 y + x^3 + 
    \alpha x y
\\ \phantom{mmmmmmmm}\hfill{+ \alpha^2 x^2 + 2 x
} \end{array}&&
\\ \hline  11
& 0 & 2 & 0 & x y^2 + \alpha^6 y^2
& 0 & -1
&&&&&
\\ & 1 & 1 & 0 & x y^2 + \alpha^6 x y
& \alpha^2 & 0
& [
    [
        1,
        \alpha^6
    ],
    [
        0,
        1
    ]
]
& x y + \alpha^6 y + \alpha^6 x + 2
&\begin{array}[t]{l} \alpha^5 x^4 + y
\\       \end{array}&&
\\ & 2 & 0 & 1 & y^2 + \alpha^7 x y + \alpha^5 x + \alpha
& 0 & -1
&&&&&
\\ \hline  12
& 0 & 0 & 3 & x + \alpha^6
& 0 & -3
&&&&&
\\ & 1 & 2 & -1 & x^5 + \alpha^6 x^4 + \alpha^6 x y^2 + 2 y^2 + 2 x y + 
    \alpha^2 y
& \alpha^6 & 0
& [
    [
        1,
        \alpha^2
    ],
    [
        0,
        1
    ]
]
& x y + \alpha^6 y + \alpha^6 x + 2
&\begin{array}[t]{l} \alpha^5 x^4 + y + \alpha^6
\\       \end{array}&&
\\ & 2 & 1 & 0 & x^4 + \alpha^7 x y^2 + \alpha^5 x y + \alpha^7 y
& \alpha^6 & 0
& [
    [
        1,
        \alpha^2
    ],
    [
        0,
        1
    ]
]
& y^2 + \alpha^7 x y + \alpha^2 y + \alpha^5 x + \alpha^2
&\begin{array}[t]{l} \alpha^5 x^3 y + 2 x^4 + \alpha^7 x^2 y + 
    \alpha x^3 + \alpha x y
\\ \phantom{mmmmmmmm}\hfill{+ \alpha^3 x^2 + \alpha^5 x
} \end{array}&&
\\ \hline  13
& 0 & 1 & 2 & x y + \alpha^6 y
& 0 & -2
&&&&&
\\ & 1 & 0 & 2 & x y + \alpha^6 y + \alpha^6 x + 2
& 0 & -2
&&&&&
\\ & 2 & 2 & -1 & x^4 y + \alpha^7 x^5 + \alpha^2 x^4 + \alpha^5 x y^2 + \alpha y^2 + \alpha^3 x y +
    \alpha^6 y
& 2 & 0
& [
    [
        1,
        1
    ],
    [
        0,
        1
    ]
]
& y^2 + \alpha^7 x y + \alpha^2 y + \alpha^5 x + \alpha^2
&\begin{array}[t]{l} \alpha^5 x^3 y + 2 x^4 + \alpha^7 x^2 y + \alpha x^3 + \alpha x y
\\ \phantom{mmmmmmmm}\hfill{+ \alpha^3 x^2 + \alpha^5 x + 2
} \end{array}&&
\\ \hline\end{array}$}}}

\caption{
Steps for correcting two errors in positions on a vertical line.
}
\end{table}

\section{The key equation for one-point codes}
\label{s:one point}

Sakata's generalization of the Berlekamp-Massey algorithm
was originally designed for a monomial ordering on a polynomial ring in
several variables~\cite{Sak90}. 
It has been adapted to the more general setting of a ring with an
order function \cite{Hoholdt:Chap}, which corresponds to an 
algebraic variety (curve, surface or higher dimensional object) and a
choice of valuation on the variety~\cite{OS:new}.   
In the case of a curve $\cal{C}$, one takes
the ring $R$ of functions having poles only at a single point $Q$ on
$\cal{C}$, and  the pole order function.  
The one-point codes defined by $\cal{C}$ and
$Q$ are obtained by evaluating functions in $R$ at rational points
$P_1,P_2,\dots,P_n$ that are distinct from $Q$.

In this section we show that the results in the Hermitian codes
section, with very minor modifications, apply to one-point codes.
The main challenge is to establish the dual bases in which we write
the locator polynomial and the evaluator, which is now a differential.  
Once this foundation is set, the decoding material falls in
place via the same arguments as were used for Hermitian codes. 
We simply state the results here and leave verification to the reader.
The section starts with  a quick tour of the main properties of
uniformizing parameters, differentials, residues, and other topics
that are needed to establish the algorithms and formulas for decoding.
Our primary reference for this section is Stichtenoth's
book~\cite{Stich:Book}, but another valuable resource is 
Pretzel's book~\cite{Pretzel:Book}.

\subsection{Curves, function fields and differentials}
Let $K$ be a function field of transcendence degree one over $\F_q$.
Let $\Cc $ be the smooth  curve over $\F_q$ defined by $K$.
We  assume that $\F_q$ is algebraically closed in $K$, which is
equivalent to $\Cc $ being absolutely irreducible.
Let $Q$ be a rational point of $\Cc $ and let $\val_Q$ be the associated
valuation of $K$. 
Let $L(mQ) $ be the space of functions on $\Cc $ having poles only at $Q$ and
of order at most $m$ there.  
Each $L(mQ)$ contains  $L((m-1)Q)$, and is either equal to it, when we
say  $m$ is a {\it gap}, or of dimension one larger, when $m$ is a
{\it nongap}.  
Let $\Lam$ be the set of nongaps and let $\Lamc$ be its complement in $\Z$. 
$\Lam$ is called the Weierstrass semigroup of $\Cc$ at $Q$.
The union of the $L(mQ) $ is a ring,
$$R= \bigcup_{m=0} ^\infty L(mQ)$$
For $f \in R$, we define $\rho(f)=-\val_Q(f)$ to be the pole
order of $f$ at $Q$.  Formally, we set $\rho(0) = -\infty$.

Let $\kap$ be the smallest positive element of $\Lam$.  
For each $b=0,\dots,\kap-1$, let $\lam_b$ be the smallest element of $\Lam$ 
congruent to $b$ modulo $\kap$.   Any integer may be written in a 
unique way as $\lam_b+ a \kap $ for some $ b \in \{0,\dots,\kap-1\}$
and $a\in \Z$. Elements of $\Lam$ have $a \geq 0$ and elements of
$\Lamc$ have $a<0$.
The set $\lambda_1,\dots,\lambda_{\kappa-1}$
is usually known as the {\it Ap\'ery} set of $\Lambda$ (named so after \cite{apery}).
Let $x \in R$ have pole order $\kap$, and for each $b$, let 
$z_b$ have pole order $\lam_b$.  
We also assume that some uniformizing parameter $u_Q$ at $Q$ has been
selected, and that $x$, and $z_b$ are  {\em monic} with
respect to $u_Q$.
That is, when either $x$ or $z_b$ is written as a power series in $u_Q$
the initial term has coefficient~1.   In particular, $z_0=1$.

\begin{proposition}
\label{p:basis}
With the notation above, $R$ is a free module over $\F_q[x]$ with
basis  $\{z_b\}_{b=0}^{\kap-1} $.  This  is also  a basis for $K $
over $\F_q(x)$. 
\end{proposition} 

\begin{proof}
Let $y \in R$ satisfy $\rho(y) \equiv b \mod \kap$.  Since $\lam_b$ is
the smallest element of $\Lam$ congruent to $b$, there is some
nonnegative $a$ such that $\rho(y)= \lam_b+a\kap$.  
 Now $\rho(y)=
\rho(x^az_b)$ so there is some $\beta \in \F_q$ such that 
$\rho(y-\beta x^az_b) <\rho(y)$.  Continuing in this manner, 
we find that for some 
$g_j \in \F_q[x]$, the pole order of $y - \sum_j g_j z_j$ is negative.
Since $y - \sum_j g_j z_j \in R$, the pole order must be $-\infty$; that is
$y-\sum_j g_j z_j=0$.  

On the other hand, no nontrivial combination $ \sum_j g_j z_j $ can equal~0.
If  $g_j\not=0$ then $\rho(g_j z_j) \equiv j \mod \kap$.
Thus $\rho( \sum_j g_j z_j) = \max_{j: g_j\ne 0}\{\rho( g_jz_j)\}$ which is
not $-\infty$.  Thus, $R$ is free over $\F_q[x]$ with basis
$\{z_j\}_{j=0}^{\kap-1}$.
The argument for linear independence  holds for $g_j \in \F_q(x)$ as
well.
Since $x$ has only one pole, and that of order $\kap$, the dimension
of $K$ over $ F_q(x)$ is $\kap$,  \cite[I.4.11]{Stich:Book}.
Thus $\{z_j\}_{j=0}^{\kap-1}$ is a basis for $K$ over $\F_q(x)$.
\end{proof}

There are parallel constructions for differentials.  
The module of differentials of $K$ over $\F_q$, which we denote $\Om$,
is a one-dimensional vector space over $K$.  
For any separating element  $ u \in K$, in particular for a
uniformizing parameter,  $du$ is a basis for $\Om$.  
If $u_P$ is  a uniformizing parameter at a point $P$, then 
any $\om \in \Om$  may be written in the form $\sum_{i=r}^{\infty} c_i
u_P^i du_P$ with $c_i\in \F_q$ and  $c_r \not=0$.  
One defines $\val_P(\om)= r$ and $\ress_P (\om) = c_{-1}$
(or $\ress_P(\om)=0$ if $r >-1$).
These definitions are independent of the choice of uniformizing parameter.   
We will say that $\om$ is {\it monic}, relative to $u_P$, when 
$c_r =1$.  
The divisor of $\om$ is $(\om) = \sum_P
\val_P(\om)$, where the sum is over all points of $\Cc$.
For any divisor $D$, $\Om(D)$ is the space of differentials such that
$(\om) \geq D$.  
Thus, $\Om(mQ)$ is the space of differentials which have valuation {\it
at least} $m$ at $Q$ and which have nonnegative valuation elsewhere.
Let 
$$\Om(-\infty Q) = \bigcup_{m=0} ^{\infty} \Om(-mQ)$$
It is evident that $ \Om(-\infty Q) $ is a module over $R$.  

The most fundamental invariant of the curve $\Cc$  is its genus, $g$.
We will use the following fundamental results about divisors and the genus.
\begin{itemize}
\item The degree of any differential is $2g-2$.
\item For  the point $Q$, the number of positive gaps, 
$| \N \setminus \Lam |$, is $ g$.
\item $\Om(-\infty Q)$  is isomorphic to $R$ when $(2g-2)Q$ is a
  canonical divisor.
\item The Riemann-Roch theorem:  For any divisor $D$, 
 \[\dim L(D) - \dim\Om(D) = m+1-g.\]
\item The residue theorem: For any differential $\om$, 
  $\sum_P \ress_P(\om) = 0$, where the sum is over all points of $\Cc$.
\end{itemize}

\subsection{One-point codes and their duals}

Let  $P_1, P_2,\dots,P_n$
be distinct rational points
on $\Cc $, each different from $Q$, and let $D = P_1+P_2+\dots+P_n$.
We define the evaluation map $\ev$ as follows.
\begin{align*}
\ev: R &\longrightarrow  \F_q^n \\
f&  \longmapsto  (f(P_1), f(P_2),\dots,f(P_n))
\end{align*}
Similarly, we have the residue map
\begin{align*}
\ress: \Om(-\infty Q -D) &\longrightarrow  \F_q^n \\
\om&  \longmapsto  (\ress_{P_1}( \om), \ress_{P_2}( \om), \dots,\ress_{P_n}( \om))
\end{align*}
Restricting the evaluation map to $L(mQ)$  and the residue map to
\mbox{$\Om(mQ-D)$} we get exact sequences. 
\begin{align*}
\begin{CD}
0  @>>> L(mQ-D) @>>> L(mQ)  @>>> \F_q^n \\
0  @>>> \Om(mQ)  @>>> \Om(mQ-D)  @>>> \F_q^n
\end{CD}
\end{align*}
The image codes are
$C_L(D,mQ)=\ev(L(mQ))$ and $C_\Om(D,mQ)=\ress(\Om(mQ-D))$. 

\begin{proposition}
\label{p:OPdualcodes}
The  codes $C_L(D, mQ)$ and $C_\Om(D, mQ)$ are dual.
\end{proposition}

\begin{proof}
For $f \in L(mQ)$ and $\om \in \Om(mQ-D)$, the poles of $f\om$ are
supported on $D$.  From the residue theorem,
\[ 
\ev(f)\cdot \ress(\om)= \sum_{k=1}^n \ress_{P_k}(f\om) = -\ress_Q(f\om)=0
\]
The Riemann-Roch theorem says
\begin{align*}
\dim L(mQ)  - \dim \Om(mQ) &= m+1-g\\
\dim L(mQ-D) - \dim  \Om(mQ-D) &= m-n+1-g\\
\end{align*}
Taking the difference,
\[\left(\dim L(mQ) - \dim L(mQ-D)\right) + 
\left(\dim \Om(mQ-D) -  \dim \Om(mQ)  \right) = n
\]
Thus, the codes are of complementary dimension and are orthogonal, so
they are dual codes.
\end{proof}

One consequence of the proposition is that 
the code $C_\Om(D, -Q)$ is the whole space $\F_q^n$.  
In a later section we will identify a differential, 
$h_{P_k}dx \in \Om(-Q-D)$, whose image under $\ress$ is 1 in position
$k$ and 0 elsewhere.  The syndrome of an error vector $e$ will be 
$ \sum_{k=1}^n e_k h_{P_k} dx $.

We will consider the family of codes $C_\Om(D,mQ)$.
The check matrix is constructed by taking rows of the form
$\ev(x^{a}z_b)$ for $a\kap+\lam_b \leq m$, arranged by increasing 
pole order.   As in earlier sections, we assume $c \in C_\Om(D,mQ)$ is
sent, the vector $u\in \F_q^n$ is received, and $e= u-c$, the error
vector, has weight $t$.

\subsection{The trace and a dual basis}

We have identified a basis for $K$ over $\F_q(x)$; we now seek a dual
basis for $\Om$.  The dual basis is constructed using the intimate
relationship between differentials and the trace map  of an
extension of function fields (see \cite[II.4, IV.3]{Stich:Book}, or
\cite[13.12-13]{Pretzel:Book}).
Let $\Trr$ be the trace map from $K$ to $\F_q(x)$.
Recall that the dual basis to $\{z_b\}_{b=0}^{\kap-1} $
is the unique set of elements of $K$,
$\zstar_{0},\dots,\zstar_{\kap-1} $ such that 
$\Trr (z_{b} \zstar_{j} )$ is $1$ if $b=j$ and~0 otherwise.

We will use a result that appears as Proposition~8 in Ch.~X 
of \cite{Lang:AG}:  Let $F$ be  a separable finite extension of $k(x)$ 
and let $Q_1, \dots,  Q_r$ be the distinct points over a point $P$ of $k(x)$
Let $y$ be an element of $F$.  Then 
$$\sum_{i=1}^r \ress_{Q_i} (ydx) = \ress_P( \Trr (y)dx)$$
The theorem assumes $k$
is an algebraically closed field.  
It is also true if $k$ 
is not algebraically closed provided $P, Q_i$ are
rational points since the residues are defined for rational points and 
unchanged when one passes to the algebraic closure.   
In our case, let $\infty$ be the point on the projective line where $x$
has a pole.  On $\Cc$, $x$ will also have a pole at any point mapping to $\infty$.  
Since the only pole of $x$ is  $Q$,  the
formula says $\ress_Q (ydx) = \ress_\infty (\Trr (y) dx)$ for any $y \in K$.

\begin{proposition}
\label{p:dualbasis}
For each $b \in \{0,\dots,\kap-1\}$, \, $\zstar_bdx$ is an element of 
$ \Om(-\infty Q)$, 
$-\zstar_b dx$ is monic, relative to $u_Q$,
and $\val_Q(\zstar_b dx) = \lam_b-\kap-1$.   Additionally,
\label{p:res}
\[ \ress_Q(z_j\zstar_b x^{a} dx)  = 
\begin{cases} 
-1& \text{ when } a=-1 \text { and } j=b\\
0 & \text{ otherwise }
\end{cases}
\]
\end{proposition}

\begin{proof}
We will prove the residue formula first.  Using the formula for the
residue at $Q$ and the property of the dual basis,  
\begin{align*}
\ress_Q(z_j\zstar_b x^{a}dx)  &= \ress_\infty (x^{a}
 \Trr(z_j \zstar_b) dx ) \\
&= 
\begin{cases} 
\ress_\infty( x^{a} dx)& \text{ when } j=b  \\
0 & \text{ otherwise } 
\end{cases}
\end{align*}
For the case $j=b$, note that $u=1/x$ is a uniformizing parameter at $\infty$, and 
$x^{a} dx= u^{-a}(-u^{-2}du) = -u^{-a-2}du$.  The residue is~$-1$
when  $a=-1$ and is zero otherwise. 

Now let $j \equiv \val_Q(\zstar_b dx) + 1 \mod \kap$ and let $a$ be such that
\[
\val_Q(\zstar_b dx)  = \lam_j -1 - (a+1) \kap \]
Then
\[
\val_Q( z_j \zstar_b x^{-a-1}dx)= -\lam_j + \left(\lam_j-1-(a+1)\kap\right) +(a+1)\kap = -1 
\]
Therefore, $\ress_Q(z_j \zstar_b x^{-a-1}dx) \not= 0$.  By what we proved earlier, this can only be
true when $j=b$ and $a=0$.   Therefore, $\val_Q(\zstar_b dx)= \lam_b-1-\kap$.
 Furthermore, 
since  $\res(z_b\zstar_b x^{-1}dx) = -1$,
and $z_b$ is monic,
 $-\zstar_bdx$ is also monic (relative to $u_Q$).

Finally, we show $\zstar_b dx\in \Om(-\infty Q)$. 
From the residue formula we can see that for each $z_j$ and any 
$h_j \in \F_q[x]$,
$\ress_Q (h_j z_j \zstar_b) = 0$.  Since any element of $ R$ can be
expressed in the form $\sum_{j=0}^{\kap-1} h_j z_j$, we conclude
that $\ress_Q(f \zstar_b dx)=0$ for any $f \in R$.
Now suppose that $\zstar_b dx$ has a pole at some
point $P\ne Q$.   By the strong approximation theorem, we may choose
$f\in R$ to eliminate any other poles of $\zstar_b dx$ away from $P$
and $Q$ and
we may also ensure that $\val_P(f\zstar_b dx) = -1$.  Then
$\ress_Q(f\zstar_b dx)
=-\ress_P(f\zstar_b dx) \ne 0$, which  contradicts what was shown above.
Thus $\zstar_b dx$ can have a pole only at $Q$.  
\end{proof}

\begin{proposition}
\label{p:Om}
With the notation above, $\Om(-\infty Q)$ is a free module over $\F_q[x]$ with
basis  $\{\zstar_b dx\}_{b=0}^{\kap-1} $.  This  is also  a basis for $\Om $
over $\F_q(x)$. 
\end{proposition}

\begin{proof}
Let $l(mQ) = \dim L(mQ) $ and $i(mQ)= \dim \Om(mQ)$.
From the Riemann-Roch theorem one can show
\[ l((m-1)Q) - l(mQ) = i((m-1)Q) - i(mQ) -1
\]
If $m\in \Lam$, the left hand side is~$-1$, so $i((m-1)Q) = i(mQ)$.
Conversely, if $m \in \Lamc$ then the left hand side is 0, so $i((m-1)Q) = i(mQ)
+1$ and there is some $\om \in \Om(-\infty Q)$ such that $\val_Q(\om)=
m-1$.
Thus 
\[
\{ \val_Q(\om ) +1 :  \om \in \Om(-\infty Q)\} = \Lamc =
\bigcup_{b=0}^{\kap-1}\{ \lam_b -a\kap: a>0\}
\]

We now proceed as in Proposition~\ref{p:basis}.  Let $\om \in \Om
(-\infty Q)$
and let $i$ and $a>0 $ be such that  $\val_Q(\om) = \lam_b - a\kap-1$.
There is some $\alpha\in \F_q$ such that 
$\val_Q(\om - \alpha x^{a-1}\zstar_b dx) >  \lam_b - a\kap-1$.
Continuing in this manner, there exist $g_b \in \F_q[x]$ such that  
$\om - \sum_{b=0}^{\kap-1} g_b\zstar_b dx$ has valuation at $Q$ larger
than $(2g-2)$.   
It is also in $\Om(-\infty Q)$, so it has no poles away from $Q$.
Thus $\om -\sum_{b=0}^{\kap-1} g_b\zstar_b dx=0$, for otherwise it
would have degree greater than $(2g-2)$.
This shows any $\om\in \Om(-\infty Q)$ is a combination of $\zstar_b
dx$ with coefficients in $\F_q[x]$.  

Uniqueness and the extension to $\Om$ are  shown as in
Proposition~\ref{p:basis}. 
\end{proof}

The next result is required to derive the error evaluation formula
that is analogous to Theorem~\ref{t:Hevalformula}.

\begin{proposition}
Let $M/L$ be a finite separable field extension and let $\Trr$ be the
trace map from  $M$ to $L$. 
Let  $z_1, \dots, z_n$ be a basis for $M$ over $L$ and let 
$\zstar_1 , \dots, \zstar_n  $ be the dual basis.  Then 
\begin{align}
\label{e:z_i zstar_i}
\sum_{i=1}^n z_i\zstar_i   &= 1
\end{align}
\end{proposition}

\begin{proof}
Since $M$ is finite and separable over $L$ there is some $y\in M$ such 
that $M=L(y)$. 
We will show the result first for the basis $1, y, \dots,y^{n-1}$.
Let $F(T) \in L[T]$ be the minimal polynomial  of $y$ and
let $F'(T)$ be its formal  derivative.  Let 
\begin{align*}
C(T) &= \dfrac{F(T)}{T-y} \\
&= c_{n-1}T^{n-1}+ c_{n-2} T^{n-2}+\dots + c_1T + c_0
\end{align*}
where $c_i \in M$ and $c_{n-1}=1$.   The proof of
\cite[III.5.10]{Stich:Book} (or \cite[VI.5.5]{Lang:Algebra}) shows that the dual
basis to $1,y,y^2,\dots,y^{n-1}$ is $c_0/F'(y),\dots,
c_{n-1}/F'(y)$.  For this basis, the sum in \eqref{e:z_i zstar_i}
is 
\begin{align}
\label{e:y^i}
\sum_{i=1}^{\kap-1 } y^i\dfrac{c_i}{F'(y)} &= \dfrac{1}{F'(y)} C(y)
\end{align}

In some algebraic closure of $M$, let $y_1, y_2, \dots, y_{n-1}$ be the
roots of $F$ that are distinct from
$y$ and let $y_n=y$.  Then $C(y) = \prod_{i=1}^{n-1} (y-y_i)$.
Since $F'(T) = \sum_{i=1}^n \prod_{j\not=i} (T-y_i)$,
$F'(y) = \prod_{i=1}^{n-1} (y-y_i) = C(y)$, so the sum in
\eqref{e:y^i} is~$1$ as claimed.

Now suppose $\{z_i\}$ is another basis  
let $\{\zstar_i \}$ be its dual basis, and let $\{\ystar_i \}$ be the
dual basis to $\{y^i\}$.
Let $M$ be the change of basis matrix from the $z$-basis to the
$y$-basis:  $z_a = \sum_{i=1}^n m_{a,i} y^i$.  
The change of basis matrix $\overline{M}$ from the $\zstar$ basis to
the $\ystar$ basis is $ (M^T)^{-1}$, as the following computation shows.
\begin{align*}
\del_{a,b} =
\Trr( z_a\zstar_b) & = \Trr  \left( \sum_{i=1}^n m_{a,i} y^i \sum_{j=1}^n
\overline{m}_{b,j} \ystar_j \right) \\
& =  \sum_{i=1}^n \sum_{j=1}^n   m_{a,i} \overline{m}_{b,j}
\Trr(  y^i \ystar_j) \\
& =  \sum_{i=1}^n \sum_{j=1}^n   m_{a,i} \overline{m}_{b,i}
\end{align*}
A similar computation shows $\sum_{a=1}^n z_a\zstar_a=1$,
\begin{align*}
\sum_{a=1} ^n z_a \zstar_a &= 
\sum_{a=1}^n \sum_{i=1}^n \sum_{j=1}^n  m_{a,i} y^i \overline{m}_{a,j} \ystar_j  \\
&=  \sum_{i=1}^n \sum_{j=1}^n y^i \ystar_j 
\sum_{a=1}^n  m_{a,i}  \overline{m}_{a,j} \\
&=  \sum_{i=1}^n  y^i \ystar_i 
= 1
\end{align*}
\end{proof}

\begin{example}
A natural  generalization of Hermitian codes is the norm-trace codes,
which were studied in \cite{Geil:normtrace}.  Consider the field
extension, $\F_{q^r}/ \F_q$.  Let $N$ be the norm function and $\Trr$
the trace function for this extension.  The norm-trace curve is 
$\Trr(y) = N(x) $, that is
\[ \sum_{i=0}^{r-1} y^{q^i} = x^{\frac{q^r-1}{q-1}}
\]
In the function field of this curve, $y$ is a solution to the
polynomial $F(T) \in \F_q(x)[T]$,  
$F(T) = \sum_{i=0}^{r-1} T^{q^i} -  x^{\frac{q^r-1}{q-1}} $.
Dividing by $T-y$ and substituting $\sum_{i=0}^{r-1} y^{q^i} $ for  
$x^{\frac{q^r-1}{q-1}}$ we get 
\begin{align*}
C(t) &=\frac{1}{T-Y}\left( \sum_{i=0}^{r-1} T^{q^i} - \sum_{i=0}^{r-1} y^{q^i}\right)  \\
&=\sum_{i=0}^{r-1}( T^{q^i} - y^{q^i})/(T-y)  \\
&= \sum_{i=0}^{r-1} \sum_{j=0}^{q^i-1} T^j y^{q^i -1-j}  \\
&=  \sum_{j=0}^{q^{r-1}-1} T^j
\sum_{i=\lceil \log_q( j +1)  \rceil} ^{r-1} y^{q^i -1-j}  \\
\end{align*}
We also have $F'(T)=1$.  Thus the dual basis to $1, y, \dots,
y^{q^r-1}$ is  $\ystar_0, \dots \ystar_{q^r-1}$ where 
$\ystar_j= \sum_{i=\lceil \log_q (j+1) \rceil} ^{r-1} y^{q^i -1-j}$.
\end{example}

\subsection{Polynomials for decoding}
Define the {\em error locator ideal} of $e$ to be
\[I^e=\{f\in R :f(P_k)= 0\mbox{ for
  all } k\mbox{ with }e_k\neq0\}\]
For a point $P$, let 
\[h_P = \dfrac{1}{x-x(P) } \sum _{b=0}^{\kap-1} z_b(P)\zstar_b.\]
We define the {\em syndrome} of $e$ to be 
\[ S = \sum_{k=1}^n e_k h_{P_k}.\]
As we did with Hermitian codes, we will give three justifications for
this definition of the syndrome.
The first is that the coefficients of $S$ are the products $\ev (x^a
z_b)\cdot e$.

\begin{lemma}
Let $s_{a,b}= \sum_{k=1}^n e_k (x(P_k))^a (z(P_k))^b$.  Then 
\[ S = \dfrac{1}{x}\sum_{b=0}^{\kap -1}\sum_{a=0} ^{\infty} s_{a,b}
x^{-a}\zstar_b\]
\end{lemma}

\begin{proof} 
Writing $(x-x(P))^{-1}$ as a series in $1/x$ we have 
\begin{align}
\label{e:h_P exp}
h_P &= \dfrac{1}{x }\left( \sum_{a=0}^\infty \left(\dfrac{ x(P)}{x}\right)^a\right)
\left( \sum _{b=0}^{\kap-1} z_b(P)\zstar_b \right) 
\\ 
&= \sum _{b=0}^{\kap-1}  \sum_{a=0}^\infty  
(x(P))^a z_b(P) x^{-a}\zstar_b  \notag \\
\intertext{Thus}
S &= \dfrac{1}{x } \sum_{k=1}^n e_k  \sum _{b=0}^{\kap-1}  \sum_{a=0}^\infty  
(x(P_k))^a z_b(P_k) x^{-a}\zstar_b  \notag \\
 &= \dfrac{1}{x } \sum _{b=0}^{\kap-1}  \sum_{a=0}^\infty  x^{-a}\zstar_b 
 \sum_{k=1}^n e_k  (x(P_k))^a z_b(P_k) \notag \\
&= \dfrac{1}{x}\sum_{b=0}^{\kap -1}\sum_{a=0} ^{\infty} s_{a,b}
x^{-a}\zstar_b \label{e:S exp}
\end{align}
\end{proof}

For the next two properties of the syndrome, we first need the
following lemma.
 
\begin{lemma}
The differential $h_P dx$ has  simple  poles at $P$ and $Q$ and no
other poles.  Furthermore $\ress_Q h_Pdx = -1$, so $-h_P dx$ is monic with respect to  $u_Q$.
\end{lemma}

\begin{proof}
The valuation at $Q$ of $\frac{1}{x-x(P)}\zstar_b dx$ is $\lam_b-1$,
and this is minimal for $b=0$.  Since $z_0=1$ and $-\zstar_0 dx$ is monic, 
$\val_Q(h_p dx) = \val_Q(\frac{1}{x-x(P)}\zstar_0 dx )= -1$ and the
residue is $-1$.

Using the expansion for $h_P$ in \eqref{e:h_P exp},
\begin{align*}
\ress_Q(x^i z_j h_P dx) &= 
 \sum_{a=0}^\infty \sum _{b=0}^{\kap-1}(x(P))^a z_b(P)
\ress_Q (x^{i-a-1} z_j\zstar_b dx ) \\
&= -(x(P))^i z_j(P)
\end{align*}
Extending by the linearity of the residue map, for any $g\in R$,
$\ress_Q(gh_P dx) =- g(P)$.   

We now show that $h_Pdx$ has no pole at $P'\not=P,Q$.  Suppose the
contrary,  $h_P dx$ has a pole at some $P' \not=P,Q$.  By the strong
approximation theorem, there is some $g \in R$ such that 
$gh_P dx$ has a zero at $P$, a simple pole at $P'$ and no other poles,
except at $Q$.  Using the  residue theorem we get a contradiction,
\[ 0 = g(P) = -\ress_Q(g h_P dx) = \ress_{P'} (gh_P dx) \not= 0
\]
Similarly, we may show that the pole of $h_P dx$ at $P$ is simple.
If not, we could find a $g \in R$ with 
$\val_P(g) = -\val_P(h_P dx)-1>0$.
Again, we get a contradiction,
\[0 = g(P)= -\ress_Q(gh_Pdx) = \ress_P(gh_p dx) \not=0
\]
\end{proof}

The connection between the error locator ideal and the syndrome is
now clear.

\begin{lemma}
For $f \in R$, $f \in I^e$ if and only if $fSdx \in \Om(-\infty Q)$.
\end{lemma}

\begin{proof}
From the previous lemma, $Sdx$ has a simple pole at each $P_k$ where
$e_k$ is nonzero.
Thus $fSdx \in \Om(-\infty Q)$ if and only if $\val_{P_k} (f) \geq 1$
whenever $e_k\not=0$.  This is just saying $f \in I^e$.
\end{proof}

Finally,  we show that for $f \in I^e$, $fS$ may be used for error evaluation.

\begin{lemma}
\label{l:OPeval}
Let $P_k$ be an error position and let $u_k$ be a uniformizing
parameter at $P_k$.  If $f$ is an error locator and 
$\varphi=f S dx$, then 
\begin{align}
\label{e:eval}
e_k \dfrac{df}{du_k}(P_k) &= \dfrac{\varphi}{du_k}(P_k)
\end{align}
\end{lemma}

\begin{proof}
Since $f$ vanishes at $P_k$ we can write $f= a_1u_k + a_2 u_k^2 + \cdots$.  
Each $h_{P_k}$ has a simple pole at $P_k$ and no pole at $P_j$ for $j\ne k$, so from the definition of $S$, 
\begin{align*}
 Sdx &= \left( e_k u_k^{-1} + c_0 + c_1 u_k + \cdots \right) du_k \\
\intertext{Thus}
\frac{ fSdx}{du_k } &= e_k a_1 + \cdots 
  \intertext{On the other hand, }
\frac{df}{du_k} &= a_1 + 2a_2u_k + \cdots
\end{align*}
Evaluating the two at $P_k$ amounts to setting  $u_k=0$, which gives the result.

\end{proof}
  
To compute $e_k$ using this formula,  we need $f$ to have a simple zero at $P_k$.
The formula simplifies when $x-x(P_k)$ itself is a uniformizing parameter at $P_k$,
$e_k\frac{df}{dx}(P_k) = fS(P_k)$.

\subsection{The key equation and its solution}

The key equation and the algorithm for solving it are little changed from those 
for Hermitian codes.  
We use $\kap$ instead of $q$ in the 
indexing of $z$ and $\zstar$.  The key equation uses differentials,
not just polynomials.  The key equation for Hermitian codes can be
derived from the one in this section by dividing by $dx$, whose
divisor is $(2g-2)Q$, 
and thereby shifting the pole order by $2g-2=q^2-q-2$. 
 We will simply state the main results, and leave adaptations of the proofs in the previous section to the reader.

\begin{definition}
\label{d:OPkey}
We say that $f \in R$ and $ \varphi \in \Om(-\infty Q)$ solve the key equation for
syndrome $S$ when $fS dx=\varphi$.
We say that a nonzero $f\in R$ and $ \varphi \in \Om(-\infty Q)$ 
solve the $K$-th approximation
of the key equation 
for syndrome  $S$  when the following two conditions hold.
\begin{enumerate}
\item $\rho( fSdx-\varphi) \leq 1-K$,
\item $\varphi$,  written in the $\ast$-basis, is a sum of terms whose
order is at least $2-K$.
\end{enumerate}
We will also say that $0$ and $x^{-a-1}\zstar_bdx$, for $a<0$,
solve the $a\kap+\lam_b$ key equation.  
\end{definition}

One could also express this definition in terms of the valuation $\val_Q$, 
$f$~and~$\varphi$ solve the $K$-th key equation when $\val_Q(fSdx-\varphi) \geq K-1$.
Since each $h_Pdx$ has a simple pole at $Q$,  $\val_Q(S dx) \geq -1$.
Therefore,  $\rho(z_b S dx) \leq 1-\lam_b$, so the  pair $z_b, 0$
satisfies the $-\lam_b$ key equation.  
The pair $0, \zstar_b$ solves the $\lam_b -\kap$ key equation.

Here are the three lemmas used in the proof that the decoding
algorithm works.

\begin{lemma}
Suppose that $f \ne 0$ and that $f, \varphi$ satisfy the $K$th key
equation for syndrome $S$.
If $g$ and $h$ are both monic of order $K$ then $\ress_Q(gfS) = \ress_Q(hfS)$.
\end{lemma}

\begin{lemma}
\label{l:OPshift}
Suppose that $f, \varphi$ satisfy the $K$th key equation.  For any
nonnegative integer $i$,  $x^if,x^i\varphi$ satisfy the $K-i\kap$ key equation.
\end{lemma}

\begin{lemma}
\label{l:OPupdate}
Suppose that $f, \varphi$ and $g, \psi$ satisfy the $K$th key equation
where $K= a\kap+\lam_b$.  Suppose in addition that $f\not=0$ and
$gSdx-\psi$ is monic of  order $1-K$.  
Let the coefficient of $x^{-a-1}\zstar_b$ in $fSdx$ be $\mu$.  
Then $f-\mu g, \, \varphi-\mu \psi$ satisfy the $(K+1)$th key equation.
\end{lemma}

The decoding algorithm has only minor changes: $\kap$ replaces $q$,
$z_i$ replaces $y^i$ and $\lam_i $ replaces $i(q+1)$.

\begin{center}
{\bf Decoding algorithm for  one-point codes}
\end{center}

\noindent{\bf Initialize:}
For $i=0$ to $\kap-1$, set
$\left(\begin{array}{cc}
f_i^{(0)}&\varphi_i^{(0)}\\g_i^{(0)}&\psi_i^{(0)}\\
\end{array}\right)=
\left(\begin{array}{cc}z_i&0\\0& -\zstar_i dx\\
\end{array}\right)$

\medskip

\noindent {\bf Algorithm:} For $m=0$ to $M$, and for each pair $i,j$ such that $m\equiv i+j\mod \kap$, set

\begin{tabular}{ll}
$d_i=\rho({f_i^{(m)}})$ & $  \qquad d_j = \rho({f_j^{(m)}})$\\ \medskip

$r_i=\frac{m-d_i - \lam_j}{\kap} $&$ \qquad r_j=\frac{m-d_j -
    \lam_i}{\kap} $ \\ \medskip

$\tf_i =z_j f_i $&$ \qquad \tf_j = z_if_j$\\  \medskip

$\mu_i = \sum_{c=0}^{\kap-1} \sum_a (\tf_i)_{a,c} s_{a +r_i,c} $&$ \qquad 
\mu_j = \sum_{c=0}^{\kap-1} \sum_a (\tf_j)_{a,c} s_{a+r_j,c}$\\  \medskip

$p=\frac{d_i+d_j -m }{\kap} -1$ &
\end{tabular}

\medskip
The update for $j$ is analogous to the one for $i$ given below.\\

$U_{i}^{(m)}=\left\{\begin{array}{ll}
\left(\begin{array}{cc}
1 & -\mu_i x^p\\
0 & 1\\
\end{array}\right)& \mbox{ if }\mu_i= 0 \mbox{ or }p\geq 0
\\
\left(\begin{array}{cc}
x^{-p}&-\mu_i\\
1/\mu_i& 0
\end{array}\right)
& \mbox{ otherwise.}
\\
\end{array}
\right.$

\medskip
\begin{quote}
$\left(\begin{array}{cc}
f_i^{(m+1)}&\varphi_i^{(m+1)}\\g_j^{(m+1)}&\psi_j^{(m+1)}\\
\end{array}\right)
=U_i^{(m)}\left(\begin{array}{cc}
f_i^{(m)}&\varphi_i^{(m)}\\g_j^{(m)}&\psi_j^{(m)}\\
\end{array}\right)$
\end{quote}

\noindent {\bf Output:} $f_i^{(M+1)}, \varphi_i^{(M+1)}$ for $0 \leq i <\kap$.
 
\medskip
One can check that at iteration $m$,  $\rho(x^{r_i}z_jf_i^{(m)})=m$.
Using an argument analogous to the one in Lemma~\ref{l:t_ab}  
one can show that $\mu_i = \ress_Q ( x^{r_i} z_jf_i^{(m)} Sdx)$ and
that this is the coefficient  of $x^{-r_i-1}\zstar_j$ in $f_i^{(m)}S$.  

\begin{theorem}
\label{t:OPBMS}
For $m\geq 0$, 
\begin{enumerate}
\item
$f_i^{(m)}$ is monic and $\rho(f_i^{(m)})\equiv i \mod \kap$.
\item 
$f_i^{(m)}, \varphi_i^{(m)}$ satisfy the $m-\rho(f_i^{(m)})$ approximation of the key equation. 
\item
$g_i^{(m)}, \psi_i^{(m)} $  satisfy the $\rho(f_i^{(m)}) -\kap$
approximation of the key equation and $g_iS dx-\psi_i^{(m)}$ is monic
of order $1 +\kap-\rho(f_i^{(m)})$.
\item
$\rho{(g_i^{(m)})} <  m-\rho{(f_i^{(m)})}+\kap$.
\end{enumerate}
\end{theorem}

The iteration at which the algorithm can terminate depends on the set
$\Delta^e=\Lambda-\rho(I^e)$ and the values $\sigma_i=\min\{\rho(f): f \in I^e
\text{ and } \rho(f) \equiv i \mod \kappa\}$. 
 
\begin{proposition}
\label{p:OPconverse}
If $\ress_Q(x^a\zstar_b fS dx) = 0$ for all $a,b$ such that that
\mbox{$a\kap + \lam_b \in \Delta^e$}  then $f\in I^e$.
In particular, if $f, \varphi$ 
satisfy the $\max \Delta^e$ key equation,
then $f \in I^e$.
\end{proposition}

\begin{proposition}
Let $\sigma_{\max}=\max\{\sigma_i:0\leq i\leq \kap-1\}$
and let \mbox{ $\del_{\max}=\max\{c \in \Delta^e\}$}.
For  $m >\sigma_{\max}+\del_{\max}$, 
each of the polynomials $f_i^{(m)}$ belongs to $I^e$.
Let $M = \sigma_{\max}+ \max\{ \del_{\max}, 2g-1\}$.
Each of the pairs $f_i^{(M+1)}, \varphi_i^{(M+1)}$ satisfies the key equation.
\end{proposition}

\subsection{Error evaluation without the error evaluator polynomials}

The error evaluation formula that we derived for Hermitian codes
carries over to one-point codes.
We have to stipulate that $x-x(P_k)$ has a simple zero at $P_k$,
though it may be possible to remove this restriction.   
As was mentioned in the section on Hermitian codes, the derivation of
the formula depends on   the fact that at iteration $m$, and for $i+j
\equiv m \mod \kap$,  $\mu_i=\mu_j$ in the decoding algorithm. 
This is proven in Proposition~\ref{p:OPdeterminants} below.

 In the proof of the proposition we will use the Cauchy-Binet Theorem.  Let
$\bB, \bC$ be $n\times 2$ matrices and let $\bT$ be an $n\times n$
matrix such that $\bC= \bT\bB$.  For $I, J$ two-element subsets of
$\{1,\dots,n\}$, let $\bC_I$ be the two rows of $\bC$ indexed by $I$
and let $\bT_I^J$ be the $2\times 2$ submatrix of $\bT$ consisting of
entries from the rows in $I$ and the columns in $J$.
The Cauchy-Binet theorem says that  
\[ \det \bC_I = \sum_J \det \bT_I^J \det \bB_J \]
where the sum runs over all two-element subsets $J$ of $\{1,\dots,n\}$.

\begin{proposition}
\label{p:OPdeterminants}
In  the $m$th iteration of the algorithm, 
$\mu_i=\mu_j$ for  $i+j\equiv m \mod \kappa$.
Furthermore for  $i\ne j$, 
the coefficient of $\zstar_0$ in the $\star$-basis expansion of each
of the following determinants is $0$:
\[ \det \begin{pmatrix} f_i ^{(m)} & \varphi_i ^{(m)}\\
 f_j ^{(m)} & \varphi_j ^{(m)}\end{pmatrix} ,
\qquad
 \det \begin{pmatrix} g_i ^{(m)} & \psi_i ^{(m)}\\
 g_j ^{(m)} & \psi_j ^{(m)}\end{pmatrix} ,
\qquad
 \det \begin{pmatrix} f_i ^{(m)} & \varphi_i ^{(m)}\\
 g_j ^{(m)} & \psi_j ^{(m)}\end{pmatrix}. \]
The coefficient of $\zstar_0$ is $-dx$ in 
\begin{align}
\label{e:res}
 \det \begin{pmatrix} f_i ^{(m)} & \varphi_i ^{(m)}\\
 g_i ^{(m)} & \psi_i ^{(m)}\end{pmatrix}.
\end{align}
\end{proposition}

The formulas  may also be expressed using residues via
  Proposition~\ref{p:dualbasis}.   
The coefficient of $\zstar_0$ in $D$ is 0 if and only if 
$ \ress_Q\left(x^{a} D \right) = 0$ for all $a$.
Equation~\eqref{e:res}
is equivalent to saying that 
\[ \ress_Q\left(x^{a} \det \begin{pmatrix} f_i ^{(m)} & \varphi_i ^{(m)}\\
 g_i ^{(m)} & \psi_i ^{(m)}\end{pmatrix} \right) = \begin{cases}
 1 & \text{ if } a = -1 \\
 0 & \text{ otherwise}
 \end{cases}.
\]

\begin{proof}
The proof proceeds by induction.  The determinental conditions are readily verified for $m=0$.  The inductive step has two parts.  First, we show that if the determinental conditions hold for $m$, then 
$\mu_i=\mu_j$ for $i+ j \equiv m \mod \kap$ in the  $m$th iteration of the algorithm.  Then we show that the determinental conditions hold for $m+1$.

Assume the determinental conditions hold for $m$.
Let $i+j \equiv m \mod \kap$ and let $\mu_i$, $\mu_j$, $r_i$, and $r_j$ be as in the algorithm.
We will suppress the superscript $(m)$ on $f_i^{(m)}$ and the other data.
We will show below that 
\begin{align}
\label{e:resform1}
\mu_i &= \ress_Q \left( x^{-p-1}f_jf_iS dx - x^{-p-1}f_j\varphi_i \right). \\
\intertext{One of the hypotheses of the lemma is that the coefficient
  of $\zstar_0$ 
in $f_j\varphi_i - f_i\varphi_j$ is 0.  Thus, we may substitute $f_i\varphi_j$ for $f_j\varphi_i$ in 
\eqref{e:resform1} to say that }
\mu_i &= \ress_Q \left( x^{-p-1}f_jf_iS dx - x^{-p-1}f_j\varphi_i \right). 
\end{align}
The right hand  side of this  formula is the analogue of \eqref{e:resform1} with $j$ and
$i$ switched.  This shows that $\mu_j=\mu_i$.

To establish \eqref{e:resform1}, 
we apply item~(2)  of  Theorem~\ref{t:OPBMS} to obtain  
\[ \rho{(f_iS dx-\varphi_i)} \leq 1+\rho({f_i}) -m.
\]
As we noted before  Theorem~\ref{t:OPBMS}, 
$\mu_i $ is the coefficient
 of $x^{-r_i-1}\zstar_j$ in $f_iS$.  
Thus,
\[ \rho{(f_iSdx-\varphi_i - \mu_i x^{-r_i-1}\zstar_j dx)}
< 1+\rho({f_i}) -m.
\]
Multiplying by $x^{-p-1 } f_j$ we have 
\[ \rho{(x^{-p-1}f_jf_iSdx-x^{-p-1}f_j\varphi_i - 
\mu_i   x^{-p-1}f_jx^{-r_i-1}\zstar_j dx})  < 1.
\]
Equivalently, the valuation of the expression is nonnegative.
This shows that 
$\ress_Q(x^{-p-1}f_jf_iSdx-x^{-p-1}f_j\varphi_i )
= \ress_Q(\mu_i   x^{-p-r_i-2}f_j\zstar_j dx)$.  The expression on the right has valuation $-1$,
and residue $\mu_i$, which establishes \eqref{e:resform1}.

We now prove that the determinental conditions of the lemma hold for
$m+1$. 
Let 
\[B_i^{(m)} = 
\begin{pmatrix} 
f_i^{(m)}&\varphi_i^{(m)}\\g_i^{(m)}&\psi_i^{(m)}\\
\end{pmatrix}, \qquad \text{ and let } \quad
\bB^{(m)} = \begin{pmatrix}
B_0^{(m)} \\
B_1^{(m)} \\
B_2^{(m)} \\
\dots \\
B_{\kap-1}^{(m)} 
\end{pmatrix}
\]
 Let $\bT$ be the update matrix for the $m$th iteration, 
so $\bB^{(m+1)} = \bT \bB^{(m)}$.
We want to show that for $I \subseteq \{1,\dots, 2\kap\}$ and $\bB_I^{(m+1)}$
the appropriate $2\times 2$ submatrix, the coefficient of $\zstar_0$
in $\det \bB_I^{(m+1)}$ is 0 unless $I$ is a consecutive pair of the
form  $\{2i+1, 2i+2\}$ for $i=0,\dots,\kap-1$.  From the inductive
hypotheses, the coefficient of $\zstar_0$ in $\det \bB_I^{(m)}$ is 
only nonzero for these  $I$.  Consequently, from the Cauchy-Binet theorem
\begin{align}
\label{e:res2}
 \ress_Q \left(x^a \det \bB^{(m+1)}_I \right)= \sum_J \ress_Q \left(x^a\det \bT_I^J \det \bB_J^{(m)} \right) 
 \end{align}
where the sum runs over all $J$  of the form 
$\{2j+1,2j+2\}$.

From the algorithm, for $i+j\equiv m \mod \kap$ and $i\ne j$,
\begin{align}\label{e:Hupdate}
\begin{pmatrix}
f_i^{(m+1)} & \varphi_i^{(m+1)}\\
g_i^{(m+1)} & \psi_i^{(m+1)}\\
f_j^{(m+1)} & \varphi_j^{(m+1)}\\
g_j^{(m+1)} & \psi_j^{(m+1)}\\
\end{pmatrix}
=
\begin{cases} 
\begin{pmatrix}
1 & 0 & 0 & -\mu x^p\\
0 & 1 & 0 & 0 \\
0 & -\mu x^p & 1 & 0\\
0 & 0 & 0 & 1 \\
\end{pmatrix}
\begin{pmatrix}
f_i^{(m)} & \varphi_i^{(m)}\\
g_i^{(m)} & \psi_i^{(m)}\\
f_j^{(m)} & \varphi_j^{(m)}\\
g_j^{(m)} & \psi_j^{(m)}\\
\end{pmatrix},
& \mbox{ if }\mu= 0 \mbox{ or }p\geq 0
\\
 \begin{pmatrix}
 x^{-p} & 0 & 0 & -\mu\\
 0 & 0 & 1/\mu & 0 \\
 0 & -\mu & x^{-p} & 0\\
 1/\mu & 0 & 0 & 0 \\
 \end{pmatrix}
 \begin{pmatrix}
 f_i^{(m)} & \varphi_i^{(m)}\\
 g_i^{(m)} & \psi_i^{(m)}\\
 f_j^{(m)} & \varphi_j^{(m)}\\
 g_j^{(m)} & \psi_j^{(m)}\\
 \end{pmatrix}
& \mbox{ otherwise.}
\end{cases}
\end{align}

Notice that we have used $\mu=\mu_i=\mu_j$.
Of course, if $i=j$, {\em i.e.} $2i=m \mod \kap$, then the formula is
simpler, $B_i^{(m+1)}= U_i^{(m)}B_i^{(m)}$, with $U_i^{(m)}$ from the algorithm.

In the formula \eqref{e:res2},
we consider two cases for $I$.  If there is no $i, j$ with  $i+j \equiv m \mod \kap$ 
such that $I \subseteq \{ 2j+1, 2j+2, 2i+1, 2i+2\}$ then for all 
$J$  of the form  $\{2j+1,2j+2\}$, $\bT_I^J$ has a row that is all
zeros, and $\det \bT_I^J = 0$.  For such $I$,
we therefore have $\ress_Q\left(x^a\det \bB_I^{(m+1)}\right)=0$.

Now consider  $I \subseteq \{ 2j+1, 2j+2, 2i+1, 2i+2\}$ 
with $i+j\equiv m \mod \kap$.  Similar reasoning shows that 
\[\ress_Q\left(x^a  \det \bB^{(m+1)}_I\right) = \ress_Q\left( x^a\det \bT_I^J \det \bB_J^{(m)}\right) +
\ress_Q\left( x^a\det \bT_I^{\bar{J}} \det \bB_{\bar{J}}^{(m)} \right)\] 
where $J= \{2j+1,2j+2\}$ and  $\bar{J} = \{2i+1, 2i+2\}$.
There are  $\binom{4}{2}$ choices of $I$ to check for each of the two
possible update matrices.
If $I=J$ or $I=\bar{J}$, then either 
$\det \bT_I^J=1$ and $\det \bT_I^J=0$ or vice-versa depending on the 
matrix.  Thus the induction hypothesis shows that the coefficient of
$\zstar_0$ in $\det \bB_I^{(m)}$ is $-dx$ as desired.
For $I=\{2i+1, 2j+2\}$ or $\{2i+2, 2j+2\}$, 
and for either update matrix,  $\det \bT_I^J= \det \bT_I^{\bar{J}}=0$.
Thus the coefficient of
$\zstar_0$ in $\det \bB_I^{(m)}$ is $0$ as desired.
Finally, for $I= \{2i+1, 2j+1\}$, and for either update matrix,
$\det \bT_I^J  = -\det \bT_I^{\bar{J}}$ and this is a monomial in $x$.  
\[ \ress_Q\left( x^a \det \bB^{(m+1)}_I \right)= \ress_Q\left(x^a\det
\bT_I^J
\left( \det \bB_J^{(m)} - \det \bB_{\bar{J}}^{(m)}\right) \right) \]  
The induction hypothesis
says that the coefficient of $\zstar_0$ is the same 
in  $\det \bB_J^{(m)}$ and $ \det \bB_{\bar{J}}^{(m)}$.
Thus the coefficient of $\zstar_0$ in $ \det \bB^{(m+1)}_I $ is 0 as desired.
\end{proof}

\begin{proposition}
\label{p:OPsum dets}
Let $B_i^{(M)} = 
\begin{pmatrix}
f_i^{(m)}  &\varphi_i^{(m)}  \\
g_i^{(m)}  &\psi_i^{(m)}  \\
\end{pmatrix}$.  Then for all $m$, 
\begin{equation}
\sum_{i=0}^{\kap -1} \det B_i^{(m)} = - dx \left(\sum_{i=0}^{\kap -1} z_i \zstar_i\right)
=-dx
\end{equation}
\end{proposition}

\begin{theorem}
\label{t:OPevalformula}
Suppose that  $x-x(P_k)$ is a uniformizing parameter at an error position  $P_k$.
Let $f'= df/dx$. Then 
\begin{align}
e_k &=  \left(\sum_{i=0}^{\kap-1} f_i'(P_k)g_i(P_k) \right)^{-1}
\end{align}
\end{theorem}

\section{Bibliographical notes}
The history of the key equation may be divided into three
stages.  In the first stage there is the key equation and
iterative solution of it in Berlekamp's book~\cite{Berlekamp:Book},
and a more implementation oriented approach in Massey's article~\cite{Massey:Shift}.  
These articles build on the Peterson-Gorenstein-Zierler decoding
algorithm~\cite{Peterson,GorZ} and Forney's
improvements~\cite{Forney:BCH}, which use matrices and are less
efficient.  

The second stage includes two new algorithms.
Sugiyama et al~\cite{sugiyama:key} define a key
equation and give an efficient solution to it  using the Euclidean
algorithm. The Welch-Berlekamp
algorithm~\cite{WelchBerl:patent} is related to the rational
interpolation problem and has  its
own key equation.
A number of articles explore the algebraic formulation of these
algorithms, efficient implementation, or the relationship between
the different algorithms.  Among these we mention 
Fitzpatrick's article on the key equation \cite{Fitz:Key},
comparisons of the Euclidean and Berlekamp-Massey algorithms by
Dornstetter~\cite{Dornstetter} and Heydtmann and Jensen
\cite{HeyJensen:BMandE}, and comparisons of key equations 
in Moon and Gunther\cite{Moon},  Morii
and Kasahara~\cite{Morii:Key}, and Yaghoobian and Blake \cite{YagBlake}.
A more extensive discussion and bibliography may be found in
Roth's textbook~\cite[Ch. 6]{Roth:Book}.

A third stage concerns the extension of the key
equation and decoding algorithms to algebraic geometry codes.
The key breakthrough was Sakata's algorithm for finding linear
recurrence relations for higher dimensional arrays~\cite{Sak90}.
We are using K\"otter's version of the algorithm for algebraic
curves~\cite{kot98},
in which the ring of functions is treated as a module over a
polynomial ring.
The Forney formula is generalized for one-point codes in
Hansen et al~\cite{Hansen}  and in
Leonard~\cite{Leonard:Forney,Leonard:Efficient}. 
Several generalizations of the key equation have appeared.
Chabanne and  Norton~\cite{ChabNorton:Key} work with a polynomial ring in
several variables and express the syndrome as a power series.
The key equation is generalized to arbitrary codes on curves 
by Ehrhard~\cite{Ehrhard:Key},
Porter, Shen and  Pellikaan \cite{PorterShen:Key},
and  by Farr{\'a}n~\cite{Farran:Key}.
A later paper by  Shen and Tzeng~\cite{Shen:Key}, deals with one-point
codes.  There are elements of all these approaches in this chapter,
but we have maintained the focus on one-point codes, where the
generalizations are particularly simple, and the treatment is based on
the articles of O'Sullivan~\cite{OS:herm,OS:key,OS:kot}.

\section*{Acknowledgment}

This work was partly supported by the Spanish Ministry of Education
through projects TSI2007-65406-C03-01 ``E-AEGIS'' and CONSOLIDER
CSD2007-00004 ``ARES'', and by the Government of Catalonia under grant
2005 SGR 00446.

\end{document}